\documentstyle[12pt]{article}
\if@twoside\oddsidemargin25mm
\else\oddsidemargin25mm\evensidemargin25mm\marginparwidth25mm\fi%
\begin{document}
\newtheorem{lemma}{Lemma}
\renewcommand{\thelemma}{\thesection.\arabic{lemma}}
\newsavebox{\uuunit}
\sbox{\uuunit}
    {\setlength{\unitlength}{0.825em}
     \begin{picture}(0.6,0.7)
        \thinlines
        \put(0,0){\line(1,0){0.5}}
        \put(0.15,0){\line(0,1){0.7}}
        \put(0.35,0){\line(0,1){0.8}}
       \multiput(0.3,0.8)(-0.04,-0.02){12}{\rule{0.5pt}{0.5pt}}
     \end {picture}}
\newcommand {\unity}{\mathord{\!\usebox{\uuunit}}}
\newcommand{\half}{{\textstyle\frac{1}{2}}}
\newcommand{\for}{{\textstyle\frac{1}{4}}}
\newcommand{\ft}[2]{{\textstyle\frac{#1}{#2}}}
\newcommand  {\Rbar} {{\mbox{\rm$\mbox{I}\!\mbox{R}$}}}
\newcommand  {\Hbar} {{\mbox{\rm$\mbox{I}\!\mbox{H}$}}}
\newcommand {\Cbar}
    {\mathord{\setlength{\unitlength}{1em}
     \begin{picture}(0.6,0.7)(-0.1,0)
        \put(-0.1,0){\rm C}
        \thicklines
        \put(0.2,0.05){\line(0,1){0.55}}
     \end {picture}}}
\newsavebox{\zzzbar}
\sbox{\zzzbar}
  {\setlength{\unitlength}{0.9em}
  \begin{picture}(0.6,0.7)
  \thinlines
  \put(0,0){\line(1,0){0.6}}
  \put(0,0.75){\line(1,0){0.575}}
  \multiput(0,0)(0.0125,0.025){30}{\rule{0.3pt}{0.3pt}}
  \multiput(0.2,0)(0.0125,0.025){30}{\rule{0.3pt}{0.3pt}}
  \put(0,0.75){\line(0,-1){0.15}}
  \put(0.015,0.75){\line(0,-1){0.1}}
  \put(0.03,0.75){\line(0,-1){0.075}}
  \put(0.045,0.75){\line(0,-1){0.05}}
  \put(0.05,0.75){\line(0,-1){0.025}}
  \put(0.6,0){\line(0,1){0.15}}
  \put(0.585,0){\line(0,1){0.1}}
  \put(0.57,0){\line(0,1){0.075}}
  \put(0.555,0){\line(0,1){0.05}}
  \put(0.55,0){\line(0,1){0.025}}
  \end{picture}}
\newcommand{\Zbar}{\mathord{\!{\usebox{\zzzbar}}}}
\newcommand{\cmap}{{\bf c} map}
\newcommand{\rmap}{{\bf r} map}
\newcommand{\crmap}{{{\bf c}$\scriptstyle\circ${\bf r} map}}
\newcommand{\Ka}{K\"ahler}
\newcommand{\cB}{{\cal B}}
\newcommand{\cM}{{\cal M}}
\newcommand{\cN}{{\cal N}}
\newcommand{\cZ}{{\cal Z}}
\renewcommand{\theequation}{\thesection.\arabic{equation}}
\renewcommand{\theenumi}{({\it\roman{enumi}})}
\newcommand{\Al}{Alekseevski\v{\i}}
\newcommand{\eqn}[1]{(\ref{#1})}
\newcommand{\QED}{{\hspace*{\fill}\rule{2mm}{2mm}\linebreak}}
\begin{titlepage}
\begin{flushright} KUL-TF-92/7\\ THU-92/19 \\ hep-th/9210068
\end{flushright}
\vfill
\begin{center}
{\large\bf Symmetry structure of special geometries }   \\
\vskip 5.mm
{\bf B. de Wit$^1$}\\
Institute for Theoretical Physics, University of Utrecht\\
Princetonplein 5, 3508 TA Utrecht, The Netherlands\\[0.3cm]

{\bf F. Vanderseypen$^2$} and {\bf A. Van Proeyen$^3$} \\
Instituut voor theoretische fysica
\\Universiteit Leuven,
B-3001 Leuven, Belgium
\end{center}
\vfill
\begin{center}
{\bf ABSTRACT}
\end{center}
\begin{quote}
Using techniques from supergravity and dimensional reduction, we
study the full isometry algebra of
K\"ahler and quaternionic manifolds with special geometry. These
two varieties are related by the so-called \cmap, which can be
understood from dimensional reduction of supergravity theories or
by changing chirality
assignments in the underlying superstring theory. An important
subclass, studied in detail, consists of the
spaces that follow from real special spaces using the so-called
\rmap. We generally
clarify the presence of `extra' symmetries emerging
from dimensional reduction and give the conditions for the
existence of `hidden' symmetries. These symmetries play a major
role in our analysis. We specify the structure of the
homogeneous special manifolds as coset spaces $G/H$. These
include all homogeneous quaternionic spaces.
\vskip 5mm \hrule width 5.cm \vskip 1.mm
{\small\small
\noindent $^1$ On leave of absence from the Institute for
Theoretical Physics, Utrecht, The Netherlands; E-mail BDEWIT at
FYS.RUU.NL\\
\noindent $^2$ Bitnet FGBDA46 at BLEKUL11\\
\noindent $^3$ Onderzoeksleider, N.F.W.O. Belgium;
Bitnet FGBDA19 at BLEKUL11}
\normalsize
\end{quote}
\begin{flushleft}
October 1992
\end{flushleft}
\end{titlepage}
\section{Introduction}

Supersymmetry is an important ingredient in many of the theories
that attempt to unify the fundamental forces between elementary
particles. At energies that are low compared to the Planck scale,
the relevant
effective theories are expected to take the form of supergravity
coupled to matter. Superstrings, which at present provide the
best candidate for a completely unified theory, allow for a huge
number of classical vacua. The ones that are considered
phenomenologically viable are compactifications of the heterotic
string on Calabi-Yau manifolds (or on the possibly more
general class of $(2,2)$ superconformal theories with $c=9$) and
lead to four-dimensional theories
with $N=1$ space-time supersymmetry \cite{GrosCand}.
For such compactifications $N=2$ supergravity is relevant, as
these superconformal theories can also be used in order to
compactify type-II strings; therefore many features of these
compactifications can be understood
on the basis of $N=2$ space-time supersymmetry, irrespective of
the fact that this symmetry is not realized for the full effective
low-energy theory corresponding to the heterotic string
\cite{Seiberg,CecFerGir,DixKapLou}. In this context the coupling of
vector multiplets to $N=2$ supergravity is particularly
important. Like in many supergravity theories the spinless matter
fields in these theories parametrize a non-linear sigma model. In
this case these sigma models define K\"ahler manifolds whose structure
is encoded in a single holomorphic and homogeneous
function \cite{dWVP}. The underlying geometry of such manifolds is
called {\em special} \cite{FerStro,special}. When the
effective supergravity theory that emerges from superstring
compactifications has spinless fields whose potential is
flat, their unconstrained vacuum-expectation values
parametrize possible superstring ground states. Therefore the
moduli space of $(2,2)$ superconformal field theories with $c=9$
(and likewise the moduli spaces of Calabi-Yau three-folds)
exhibits special geometry, and this plays an
important role in the study of these manifolds and intriguing
phenomena such as mirror manifolds and generalized target-space
duality (see, for instance, \cite{FerStro,Cand,mirror,FerLusThe}
and references therein).

Special geometry is not confined to K\"ahler spaces. In $N=2$
supergravity one can also have scalar supermultiplets, consisting
of spin-0 and spin-$\half$ fields. In this case the spinless
fields lead to quaternionic non-linear sigma models \cite{BagWit}.
There is a subclass of such models whose structure is again encoded
in a single holomorphic and homogeneous function. These spaces
emerge also in $(2,2)$ compactifications of
type-II superstrings and are called {\em special} quaternionic
spaces.  In this way both a special K\"ahler and a
special quaternionic manifold are assigned to the same function.
When comparing the compactification of the type-IIA superstring
to the compactification on the same
superconformal system of the type-IIB superstring, one discovers
that the functions characterizing the K\"ahler and the quaternionic
space are interchanged \cite{Seiberg}. Hence there is a
mapping, the so-called \cmap, that is induced by changing the
superstring from type IIA to IIB, and vice versa
\cite{CecFerGir}. Within the context of
supergravity alone, it turns out that the \cmap\ is induced by
dimensional reduction of $d=4$ supergravity coupled to
vector multiplets. The resulting $d=3$ supergravity theory
couples only to quaternionic non-linear sigma models.
Supersymmetry, which is preserved by dimensional reduction,
thus ensures that every special K\"ahler space is mapped to a
quaternionic space. Note, however, that there exist quaternionic
spaces that couple to $N=2$ supergravity, but are not
{\em special}, i.e., they are not in the image of the \cmap.

In this paper we study the isometry structure of manifolds with
special geometry. Because of supersymmetry, the isometry
transformations usually define invariances of the full supergravity
theories, provided they are suitably extended to act on vector
and spinor fields. In $d=4$ dimensions such transformations
often take the form of generalized duality invariances, which
leave the field equations, but not the Lagrangian, invariant
\cite{dual}. It turns out to be convenient to introduce also a
real version of special manifolds, corresponding to the
non-linear sigma models that one obtains in $d=5$ space-time
dimensions when coupling $N=2$ supergravity to vector multiplets
\cite{GuSiTo}. Upon reduction to $d=4$ space-time
dimensions, these theories give rise to special K\"ahler
manifolds, so that one can again consider a mapping, but now from
the special real to the special K\"ahler manifolds. This leads to
the definition of the \rmap. Not all special K\"ahler manifolds
are contained in the image of the \rmap. (Following \cite{sssl} we
sometimes call the special real spaces and their \Ka\ and
quaternionic counterparts, `very special' spaces.) We thus encounter the
following relation between the real, K\"ahler and quaternionic
special spaces,
\begin{equation}
\Rbar_{n-1} \stackrel{\bf r}{\longrightarrow} \Cbar_n \ ,\qquad
\Cbar_n \stackrel{\bf c}{\longrightarrow} \Hbar_{n+1}\ ,
\end{equation}
where $n-1$, $n$ and $n+1$ denote the real, complex and
quaternionic dimension of the real, K\"ahler and quaternionic
spaces, respectively.
The special K\"ahler spaces in the image of the \rmap\ were
already studied in \cite{BEC}, because they allow supergravity
couplings with flat potentials, and are determined by a symmetric
three-index tensor $d_{ABC}$. Under mild assumptions one can show
that the corresponding couplings of vector multiplets arise from
compactifications of type-IIA superstring theories. Furthermore
nearly all homogeneous special \Ka\ and quaternionic spaces are
in this subclass \cite{Cecotti,dWVP3}. For a preliminary account
of our results, see \cite{sbstring}.

The fact that the {\bf r} and the \cmap\ are induced by
dimensional reduction has intriguing implications for the
isometry structure of the special geometries that are in the
image of (one of) these maps. First of all, the dimension of the
non-linear sigma models is increased by spinless fields that
originate from the higher-dimensional tensor and vector fields.
Secondly, part of the gauge and general-coordinate
transformations pertaining to the extra space-time
coordinate(s) remain as symmetries after reduction and lead to an
enlargement of the isometry group of the non-linear sigma model.
An extra feature is present in
the dimensional reduction to three dimensions, where abelian
vector fields are converted into scalars, which are subject
to again extra symmetries.
A more detailed analysis shows that,
upon application of the map, the rank of the space
is increased by precisely one unit and the
roots associated to the extra isometries are confined to a
particular half-space of the root lattice corresponding to the
full algebra of isometry transformations.
In addition there may
exist so-called {\em hidden} symmetries, whose existence
cannot be inferred directly from the higher-dimensional origin
of the theory (note that we deviate here from the
terminology used in the supergravity literature, where both
``extra" and ``hidden" symmetries are called hidden). The roots
corresponding to these
hidden symmetries also take characteristic positions in the root
lattice. We already presented an analysis of the isometry
structure of the special quaternionic manifolds in \cite{dWVP2};
the results for the special \Ka\ spaces in the image of the
\rmap\ follow from the analysis presented in \cite{BEC}. Some
characteristic features of the extra symmetries were
independently noted in \cite{GalOg}.

The possible existence of the hidden symmetries is analyzed in
detail, first for the generic special quaternionic spaces, and
subsequently for the \Ka\ and quaternionic spaces that are in the
image of the \rmap\ and the \crmap. The homogeneous spaces play a
special role, because under certain conditions the
homogeneity of the space is preserved by the $\mbox{\bf r}$ and the
\cmap. For homogeneous spaces the isometries act transitively on
the manifold so that every two points are related by an element
of the isometry group. The orbit swept out by the action of the isometry
group $G$ from any given point is (locally) isomorphic to the
coset space $G/H$, where $H$ is the isotropy group of that point.
For non-compact homogeneous spaces where $H$ is the maximal compact
subgroup of $G$, there exists a solvable subgroup that acts
transitively, whose dimension is equal to the dimension of the
space. Such spaces are called {\em normal}.
Some time ago a general classification of normal quaternionic
spaces\footnote{According to a conjecture \cite{Aleks} the homogeneous
quaternionic spaces consist of the normal and the compact
symmetric ones.}, i.e. quaternionic spaces that admit a solvable
transitive group of isometries, was given \cite{Aleks}.

All the normal quaternionic spaces,
with the exception of the quaternionic projective spaces, are in
the image of the \cmap, as was demonstrated explicitly in
\cite{Cecotti}. In
\cite{dWVP3} two of us gave a classification of the homogeneous
\Ka\ and quaternionic spaces that are in the image of the \crmap,
containing the above spaces as well as certain homogeneous
quaternionic and \Ka\ spaces that were overlooked in the
classification of \cite{Aleks}.
This class of spaces is characterized by realizations of real
Clifford algebras with positive-definite metric. For every
(not necessarily irreducible)
realization of a real Clifford algebra with positive signature
and arbitrary dimension, denoted by $q+1$ (so that $q\geq -1$),
there exists an homogeneous real, an homogeneous special \Ka, and an
homogeneous special quaternionic manifold.

\Al\ presented the solvable
subalgebra of the quaternionic isometry groups, which
completely determines the homogeneous spaces. The solvable algebra
does not contain the
hidden symmetries discussed above, but when the space is
symmetric, the hidden symmetries complete the solvable algebra
to a simple algebra. In this paper we determine
the hidden symmetries for the homogeneous spaces.
In this way we are able to determine the isometry group $G$
(which is not semisimple for the non-symmetric spaces) and its
compact isotropy
subgroup $H$ for all the homogeneous quaternionic spaces and the
homogeneous \Ka\ spaces in the image of the \rmap. For the real
special spaces we have similar results, but they are somewhat
less general as our methods restrict us to isometries that are
symmetries of the full $d=5$ supergravity theory containing a
non-linear sigma model with the real special space as its target space.
For the non-symmetric spaces, the Cartan subalgebra of the
isometry group $G$ contains one generator with respect to which
all the generators have eigenvalues equal to 0, 1 or 2.
Apart from this generator, the generators with eigenvalue 0
define a semisimple
group, whose maximal compact subgroup is equal to the isotropy
group  $H$. This semisimple group is ${\cal
O}\otimes {\cal S}_q$, where ${\cal O}$ equals
$SO(q+1,1)$, $SO(q+2,2)$ or $SO(q+3,3)$ for the real, \Ka,
and quaternionic case, respectively, and ${\cal S}_q$ is the
(compact)
metric-preserving group of transformations in the spinor space
that commute with the Clifford algebra. Furthermore
the generators with eigenvalue 1 form a $(spinor,vector)$ real
representation with respect to ${\cal O}\otimes {\cal S}_q$.
There are no generators with eigenvalue 2 in the real case. For
the \Ka\ case there is just one such generator, while for the
quaternionic case these generators constitute a $(vector,singlet)$
representation.

This paper is organized as follows. In section 2 we introduce
\Ka , real and quaternionic special geometry and give the
corresponding supergravity Lagrangians with particular
attention to the isometry structure of manifolds that are related
by the {\bf r} or the \cmap. We formulate the conditions for the
presence of the hidden symmetries and describe the characteristic
root-lattice decompositions that emerge upon application of one
of these maps. Section~\ref{ss:algebra} describes the main
features and some of
their consequences of the isometry algebra of the generic \Ka\ and
quaternionic special manifolds. Furthermore we analyze under
which conditions the
\cmap\ (or the \rmap) preserves the homogeneity of the
manifolds. In section~\ref{ss:exd}
we study the special real manifolds and their corresponding
\Ka\ and quaternionic spaces and exhibit their symmetry structure.
Section~\ref{ss:gamma} contains a discussion of the
subclass of these spaces that are homogeneous, deriving
the group of isometries and isotropies, described in the previous
paragraph.

Some of the more technical details have been relegated to appendices.
In appendix A we discuss some features related to the role played by
solvable algebras in the theory of non-compact homogeneous spaces.
Appendix B contains some useful formulae related to
special \Ka\ spaces.  In appendix~\ref{app:symrepar} we explain
the symplectic reparametrizations of special \Ka\ manifolds,
which encompass the so-called generalized duality invariances,
and give the transformation rules of the scalar, spinor and
vector fields.

\section{Preliminaries}   \label{ss:prel}
\setcounter{equation}{0}
In this section we introduce the special geometries that
appear in non-linear sigma models coupled to $N=2$ supergravity.
We give the relevant  supergravity Lagrangians, which,
via dimensional reduction, give rise to or originate from
$N=2$ supergravity coupled to $n$ vector
supermultiplets in $d=4$ space-time dimensions. As explained in
the introduction, dimensional reduction induces the $\bf c$
and $\bf r$ maps that relate the various types of special
geometries associated with the non-linear sigma models of the
supergravity theories. We start by summarizing the situation
for the four-dimensional theory, at least as far as the bosonic
part of its Lagrangian is concerned.  The corresponding non-linear
sigma models define the {\em special} K\"ahler
manifolds. Then we discuss how a sub-class of the $d=4$ theories
originate from supergravity in $d=5$ dimensions, whose
non-linear sigma models are based on {\em special} real
manifolds. Subsequently
we give the relation with supergravity in $d=3$ dimensions, which
lead to the {\em special} quaternionic manifolds. Throughout
this section we try to exhibit the
relation between the various non-linear sigma models with
particular emphasis on the structure of their isometry groups.

\subsection{Special K\"ahler manifolds} \label{ss:skm}
The coupling of $n$ vector multiplets to $N=2$ supergravity in
$d=4$ dimensions is encoded in a single holomorphic function $F(X)$,
which is
homogeneous of second degree in terms of the $n+1$ variables
$X^I$ labeled by indices $I=0,1,\ldots, n$. Therefore it satisfies
identities such as
$F=\frac{1}{2}F_IX^I$, $F_I= F_{IJ}X^J$, $X^IF_{IJK}=0$, where
the subscripts $I,J,\dots$ denote differentiation with respect to
$X^I$, $X^J$, etc. The bosonic Lagrangian reads \cite{dWVP,dWLVP}
\begin{eqnarray}
e^{-1}{\cal L} &=& -\textstyle{1\over2} R +
(XN\bar X)^{-1}\,\cM_{I\bar{J}}\,\partial
_\mu X^I \,\partial^\mu \bar X^J\nonumber\\
&&+\textstyle\frac{1}{4} \left\{ \cN_{IJ}\,F^{+I}_{\mu\nu}\,
F^{+\mu\nu J} +  \bar\cN_{IJ} \,F^{-I}_{\mu\nu}\,F^{-\mu\nu J} \right\} ,
\label{4dLagr}
\end{eqnarray}
where $R$ is the Ricci scalar, $F^{\pm I}_{\mu\nu}$ are the
(anti)selfdual components of
the $n+1$ field strengths and the tensors $N_{IJ}$, $\cM_{I\bar{J}}$
and $\cN_{IJ}$ are defined by
\begin{eqnarray}
N_{IJ}&=&{\frac{1}{4}}\left( F_{IJ}+\bar F_{IJ}\right),
 \nonumber \\
\cM_{I\bar J}&=&N_{IJ}-\frac{(N\bar
X)_I\,(NX)_J}{XN\bar X} , \nonumber\\
\cN_{IJ}&=&\frac{1}{4}\bar
F_{IJ}-\frac{(NX)_I\,(NX)_J}{XNX} . \label{defmat}
\end{eqnarray}
Here we used an obvious notation where $(NX)_I=
N_{IJ}X^J$, $XN\bar X=X^IN_{IJ}\bar X^J$, etc..

The Lagrangian (\ref{4dLagr}) contains $n+1$ vector fields,  with
one of them (the so-called graviphoton)
belonging to the  $N=2$ supergravity multiplet. However,
it depends on only $n$ scalar fields, because
$\cM_{I\bar J}$ has a null vector proportional to $X^I$ and $\bar
X^J$, as one easily verifies, while the tensors $\cM_{I\bar J}$,
$\cN_{IJ}$ and $N_{IJ}$ depend only on ratios of the fields
by virtue of the homogeneity of the function $F(X)$. Therefore it
is convenient to introduce  $n$ independent fields through the
ratios $z^A\equiv X^A/X^0$ ($A= 1,
\ldots,n$). These coordinates are sometimes called {\it special}
coordinates \cite{FerStro,CdAF}. Including $z^0=1$ we can
straightforwardly replace all fields $X^I$ in
the above equations by $z^I$.

The Lagrangian of these $n$ scalar fields $z^A$
takes the form of a non-linear sigma model corresponding to a
K\"ahler manifold. Its  K\"ahler potential is
\begin{equation}
K(z,\bar z ) = \ln Y(z,\bar z) = \ln N_{IJ}\,z^I \,\bar z^J,
\end{equation}
so that the sigma model metric is given by
\begin{equation}
g_{A\bar B} = \frac{\partial^2 K(z,\bar z)}{\partial z^A\,
\partial\bar z^B} =  \frac{\cM_{A\bar B}}{zN\bar z}\ ,
\label{Kmetric}
\end{equation}
where $zN\bar z\equiv z^IN_{IJ}\bar z^J$.
The curvature tensor corresponding to this metric
equals\footnote{For the precise definitions of the K\"ahler curvature
and connections, see the beginning of section~4.}
\begin{equation}
R^A{\!}_{BC}{\!}^D = -2 \delta_{(B}^A\,\delta_{C)}^D -
\frac{1}{(z N\bar z)^2}\, Q_{BCE}\,\bar Q^{EAD} ,
\label{Kcurv}
\end{equation}
where
\begin{equation}
Q_{IJK} \equiv \textstyle{\frac{1}{4}} X^0\,F_{IJK} , \ \ \ \bar
Q^{ABC} = g^{\bar D A}g^{\bar E B}g^{\bar F C} \bar Q_{DEF} .
\end{equation}
The scalar fields $z^A$ are constrained to a domain
defined by the requirement that the kinetic terms be positive.
This is the case when $Y(z,\bar z)$ is positive and $\cM_{A\bar
B}$ is negative definite (as shown in \cite{BEC} this ensures
that the matrix $\cN_{IJ} + \bar\cN_{IJ}$ is negative definite).
The {\em special} K\"ahler manifolds are thus fully specified in
terms of a homogeneous holomorphic functions of second degree.
One may also describe special geometry on the basis
of (\ref{Kcurv}) in a coordinate-independent way \cite{FerStro,CdAF}.

It is known that terms in $F(X)$ that are
quadratic polynomials in $X$ with {\em imaginary} coefficients,
do not contribute to the action. In addition, it is possible
that two different functions $F(X)$ and $\tilde F(\tilde
X)$, where the new fields $\tilde X{}^I$ can be expressed in terms
of the old ones, give rise to
the same theory in the sense that their equations of motion are
equivalent \cite{CecFerGir} (at least for abelian vector fields).
The reparametrizations that relate $X$ and $\tilde X$ are
called {\it symplectic}, because the $(2n+2)$-component vectors
$(X^I,-\half iF_I)$ and $(\tilde X^I,-\half i\tilde F_I)$ are
related by constant
$Sp(2n+2,\Rbar)$ matrices $\cal O$. Here $\tilde F_I$ is the derivative
with respect to $\tilde X{}^I$
of the new function $\tilde F(\tilde X)$, which can be determined
from the symplectic matrix and the original function $F(X)$.
The symplectic reparametrizations are discussed in more detail
in appendix~\ref{app:symrepar}. There we will also exhibit how
the other fields of
the vector supermultiplets transform under the symplectic
reparametrizations.\footnote{For Calabi-Yau manifolds these symplectic
transformations are naturally induced on the periods by
changes in the corresponding cohomology basis (see, e.g.
\cite{special,Cand}).}

When the function $F(X)$ does not change under a symplectic
reparametrization, i.e., when
\begin{equation}
\tilde F(\tilde X) = F(\tilde X)\ ,
\label{Finv2}
\end{equation}
then one has a so-called {\it duality invariance}, which
gives rise to an isometry of the non-linear sigma model. For
infinitesimal transformations, where the $Sp(2n+2,\Rbar)$ matrix
is parametrized by
\begin{equation}
{\cal O} =\unity + \left(\begin{array}{cc}
B & -D \\[1mm]
C &-B^{\rm T}
\end{array}\right) , \label{Oinf}
\end{equation}
with $B^I_{\ J}$, $C_{IJ}$ and $D^{IJ}$ real constant
$(n+1)\times (n+1)$ matrices and $C$ and $D$ symmetric,
the consistency of the symplectic
transformation and the identification \eqn{Finv2}
requires the condition \cite{dWVP}
\begin{equation}
iC_{IJ}\,X^IX^J - B^I_{\ J}\,F_I X^J -\textstyle{\frac{1}{4}} i
D^{IJ} \,F_IF_J = 0\ .
\label{masterdual}
\end{equation}
For the scalar sector this
produces invariances of the action under
\begin{equation}
\delta X^I= B^I_{\,J} \,X^J +{\textstyle{1\over 2}}iD^{IJ}\,F_J\ .
\label{scaltra}
\end{equation}
Observe that \eqn{Finv2} does not imply that the function $F(X)$
is invariant under \eqn{scaltra}; instead one finds
\begin{equation}
\delta F(X)\equiv F(\tilde X)-F(X) = i\left( C_{IJ}\,X^IX^J
+{\textstyle{1\over4}} D^{IJ}\, F_IF_J\right) \ . \label{delFd}
\end{equation}
For further details of these transformations we refer to
appendix~\ref{app:symrepar}. Finally, we should stress, that one
cannot exclude the possibility that the sigma model possesses
more isometries than just those corresponding to the above
duality transformations; these additional isometries would then
be broken by the interaction with the vector fields.

This completes our introduction to special K\"ahler
manifolds characterized by homogeneous holomorphic functions $F(X)$,
modulo quadratic polynomials with imaginary coefficients and
$Sp(2n+2,\Rbar)$ reparametrizations. Let us now briefly
discuss the class of theories corresponding to the functions
\begin{equation}
F(X)= i d_{ABC} {X^AX^BX^C\over X^0} , \label{Fd}
\end{equation}
with $d_{ABC}$ a symmetric real tensor. As we shall discuss
shortly, these theories follow from five-dimensional
supergravity by dimensional reduction, so that the sigma models
corresponding to (\ref{Fd}) constitute the image of the $\bf r$ map.
Supergravity theories based on these functions can lead to flat
potentials, as was shown in \cite{BEC}. Furthermore they
appear in the low-energy sector of certain
superstring compactifications on (2,2) superconformal theories
\cite{CecFerGir} and exhibit Peccei-Quinn-like symmetries as is
appropriate for certain (classical) superstring compactifications.
There are arguments indicating that this class of
functions comprises all the homogeneous K\"ahler spaces that can
be coupled to $N=2$ supergravity \cite{Cecotti,dWVP3}.

The functions (\ref{Fd}) lead to the following
K\"ahler metric, which depends only on
the imaginary parts of the coordinates $z^A$~,
\begin{equation}
g_{A\bar B} = \frac{\cM_{A\bar B}}{z N \bar z} = 6\,\frac{(d\,
x)_{AB}}{(d\,xxx)} - 9\, \frac{(d\,xx)_A\,(d\,xx)_B}{(d\,xxx)^2} ,
\label{scalkin}
\end{equation}
where $x^A\equiv  i(z^A-\bar z^A)$,
$(d\,x)_{AB}= d_{ABC}\,x^C$,  $(d\,xx)_A= d_{ABC}\,x^Bx^C$
and $(d\,xxx)= d_{ABC}\,x^A x^B x^C$.
Furthermore we have
\begin{equation}
Q_{ABC}= \textstyle{3\over 2} i\,d_{ABC} .
\end{equation}
The curvature corresponding to (\ref{scalkin}) follows from
\eqn{Kcurv},
\begin{equation}
R^A{}_{\!BC}{}^D = - 2\delta_{(B}^A\,\delta_{C)}^D +
\textstyle{\frac{4}{3}} C^{ADE} \,d_{BCE} \ ,\label{curvC}
\end{equation}
where $C^{ABC}$ is defined by
\begin{equation}
C^{ABC} = - \textstyle{\frac{9}{8}} i\, (zN\bar z)^{-2} \,\bar
Q^{ABC}= -27 g^{A\bar D}g^{B\bar E}g^{C\bar F}d_{DEF}(d\,xxx)^{-2}.
\label{defCABC}\end{equation}

The theory based on (\ref{Fd}) is always invariant under
duality invariances. Imposing the condition (\ref{masterdual}) we
find the following form for the matrices $B^I_{\,J}$, $C_{IJ}$
and $D^{IJ}$ (the first row and column refer
to the $I=0$ component) \cite{BEC},
\begin{equation}
B^I_{\,J} =\left(\begin{array}{cc}
\beta & a_B \\
 b^A &\tilde B^A_{\;B} +\frac{1}{3}\beta \,\delta^A_{\,B}
\end{array}\right) ,
\ \
C_{IJ} =\left(\begin{array}{cc}
0 & 0 \\
 0 & 3 d_{ABC}\,b^C
\end{array}\right) ,
\ \
D^{IJ} =\left(\begin{array}{cc}
 0 & 0 \\
 0 & D^{AB}
\end{array}\right) .    \label{matrixduald}
\end{equation}
The isometries associated with
the parameters $\beta$ and $b^A$ are always present; those
depending on $\tilde B^A_{\;B}$ correspond to the symmetries
of $d_{ABC}$, i.e.,
\begin{equation}
\tilde B^D_{\,(A}\,d_{BC)D}= 0 . \label{Binv}
\end{equation}
One can prove that the matrix $D^{AB}$ is proportional
to the parameters $a_A$,
\begin{equation}
D^{AB} = -\textstyle{\frac{4}{9}} C^{ABC}\,a_C\ ,\label{DAB}
\end{equation}
while these parameters are subject to the condition
\begin{equation}
a_{G}\, E^G_{ABCD}  = 0\ ,  \label{ainv}
\end{equation}
with the tensor $E^G_{ABCD}$ defined by
\begin{equation}
E^G_{ABCD} = C^{EFG}\, d_{E(AB}\,d_{CD)F} -  \delta^G_{(A} \,
d_{BCD)}\ .
\label{defE}
\end{equation}
Obviously the matrix $D^{AB}$ and the parameters $a^A$ must be
constant, so that $C^{ABC}a_C$ (and therefore $R^A{}_{\!BC}{}^{\!D}
a_D$) is constant. In section~\ref{ss:dKa} we will show that
\eqn{ainv} is a necessary and sufficient condition for this to be
the case.

\begin{figure}[htf]
\begin{center}
\setlength{\unitlength}{0.5mm}
\begin{picture}(150,120)
\put(20,60){\line(1,0){110}}
\put(75,10){\line(0,1){100}}
\put(75,60){\circle*{2}}
\put(75,60){\circle{5}}
\put(75,90.6){\circle*{2}}
\multiput(103.9,39.6)(0,30.6){3}{\circle*{2}}
\put(65,63){\makebox(0,0)[bl]{$\omega _1$}}
\put(65,93.6){\makebox(0,0)[bl]{$\omega _2$}}
\put(77.4,51){\makebox(0,0)[bl]{$\beta $}}
\put(106.3,36.6){\makebox(0,0)[bl]{$b^2$}}
\put(106.3,67.2){\makebox(0,0)[bl]{$b^3$}}
\put(106.3,97.8){\makebox(0,0)[bl]{$b^1$}}
\put(46.1,80.4){\makebox(0,0){$\diamond$}}
\put(43.7,77.4){\makebox(0,0)[br]{$a_2$}}
\end{picture}
\end{center}
\caption{Root lattice corresponding to the isometries of an $n=3$ special
\Ka\ manifold discussed in the text.} \label{figD}
\end{figure}
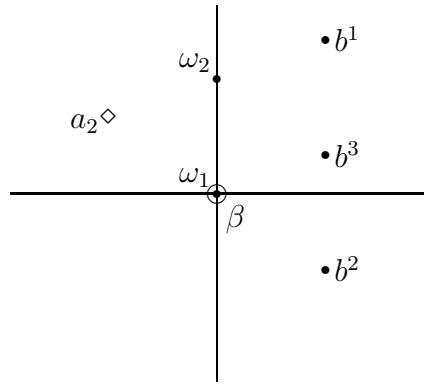

In terms of the special coordinates the above isometries imply
the following transformation
\begin{equation}
\delta z^A = b^A - \textstyle{2\over 3} \beta \,z^A +\tilde
B^A_{\;B}\, z^B + \textstyle{1\over 2} \big(R^A{}_{\!BC}{}^{\!D}\,
a_D\big)\, z^B z^C .
\label{ztrans}
\end{equation}
Clearly the transformations characterized by $\tilde B^A_{\;B}$
and $\beta$ form two commuting subgroups of the full group of
duality transformations. The root lattice of the algebra
corresponding to all
duality transformations consists of the root lattice of the
generators corresponding to $\tilde B^A_{\;B}$ extended with one
dimension associated with the eigenvalues of the roots under the
$\beta$-symmetry. This leads to a characteristic lattice such as
shown in Fig.~\ref{figD} for an $n=3$ special \Ka\ space based on
the function $F(X) =3 i\left(X^2/X^0\right)\left(X^2X^1 -
(X^3)^2\right)$.
The subgroup associated with the matrices $\tilde B^A_{\;B}$ has
two independent parameters denoted by $\omega _1$ and $\omega _2$. Their
roots correspond to the solvable algebra of $SU(1,1)$, which is
extended to a six-dimensional solvable algebra by the roots
associated with the parameters $\beta$, $b^1$, $b^2$ and $b^3$.
In this case there is one {\em hidden} symmetry associated with the
parameter $a_2$ and indicated in the diagram by $\diamond$.
For higher rank one obtains a similar lattice by projecting on a
suitably chosen plane. This particular example is discussed in
\cite{sssl} and is the simplest case of the homogeneous
non-symmetric spaces discussed in section~\ref{ss:gamma}
(it will be denoted by $L(-1,1)$).

\subsection{Special real manifolds} \label{ss:srm}
The theories corresponding to the functions (\ref{Fd}) can be
obtained by dimensional reduction from Maxwell-Einstein
supergravity theories in $d=5$ space-time dimensions \cite{GuSiTo}.
This theory contains $n-1$ real scalar fields and $n$ vector fields
(one of them corresponding to the graviphoton). The Lagrangian
corresponding to these fields reads
\begin{eqnarray}
e^{-1}{\cal L} &=& -{\textstyle{1\over 2}} R -{\textstyle{3\over
2}}\,d_{ABC} \,
h^A\,\partial_\mu h^B\,\partial^\mu h^C  \nonumber\\
&& +{\textstyle{1\over 2}}\,\left( 6(d\,h)_{AB} - 9 (d\,h h)_{A}\,
(d\,hh)_B \right) F^A_{\mu\nu}(A) \,F^{B\,\mu\nu}(A)
\nonumber\\
&& + e^{-1} i\epsilon^{\mu\nu\rho\sigma\lambda} d_{ABC} \,
F^A_{\mu\nu}(A)\,F^B_{\rho\sigma}(A)\, A^C_\lambda ,
\label{5dLagr}
\end{eqnarray}
where $A^A_\mu$ and $F_{\mu\nu}^A(A)$ denote the gauge fields and
their corresponding abelian field strengths and the scalar fields $h^A$ are
subject to the condition
\begin{equation}
d_{ABC}\,h^A h^B h^C =1.  \label{dhhh1}
\end{equation}
The scalar fields must again be restricted to a domain
so that all kinetic terms in (\ref{5dLagr}) have the required
signature.

After dimensional reduction, the Lagrangian (\ref{5dLagr}) becomes equal to
(\ref{4dLagr}); the imaginary part of the four-dimensional scalar fields
$z^A$ originate
from the components of the gauge fields $A^A_\mu$ in the fifth
dimension, while their real part corresponds to the $n-1$
independent fields $h^A$ and the component $g_{55}$
of the metric. The $n+1$ gauge fields in four dimensions are related
to the $n$ gauge fields in five dimensions and the off-diagonal
components $g_{\mu 5}$ of the metric in five dimensions.

The Lagrangian (\ref{5dLagr}) is manifestly invariant under linear
transformations of the fields
\begin{equation}
h^A \to \tilde B^A_{\;B}\, h^B, \qquad  A^A_\mu \to \tilde
B^A_{\;B}\,A^B_\mu,
\end{equation}
provided that the matrices $\tilde B$ leave the tensor $d_{ABC}$
invariant (cf. \ref{Binv}). After reduction to four space-time
dimensions a number of extra symmetries emerges, which find their
origin in the five-dimensional theory. First of all, the extra
vector field emerging from the five-dimensional metric has a
corresponding gauge invariance
related to reparametrizations of the extra fifth coordinate by
functions that depend only on the four space-time coordinates. Then
there are special gauge transformations of the $n$ vector fields
with gauge functions that depend exclusively and linearly on the
fifth coordinate. Under these transformations the fifth
component of each of the gauge fields transforms with a constant
translation, whereas the remaining four-dimensional gauge fields
transform linearly into the gauge field originating from
the five-dimensional metric (this last transformation arises
because of certain field redefinitions that must be performed on
the vector fields for reasons of $d=4$ general covariance). As only
the field equations are gauge invariant (the Lagrangian is not
gauge invariant in view of the last term in
(\ref{5dLagr})), it comes as no surprise that these invariances
will manifest themselves as duality invariances of the
four-dimensional field equations corresponding to the Lagrangian
(\ref{4dLagr}). Indeed, these transformations are the ones
associated with the parameters $b^A$.

The same phenomenon takes place for scale transformations of the
fifth coordinate, which do not leave the Lagrangian invariant
either. In the four-dimensional reduction these transformations
correspond
to the duality invariances associated with the parameter $\beta$.
However, this relationship is somewhat less direct, as it has to
be defined such that the properly defined four-dimensional metric
remains invariant under this transformation. The latter may not
be so obvious in view of the standard Weyl rescaling that is required
in order to obtain the standard Einstein-Hilbert action after
dimensional reduction. We should emphasize that none of the
extra symmetries emerging from the dimensional reduction act
on the original five-dimensional scalar fields. Of course, one
may have performed field redefinitions after the reduction that
obscure this fact.

What remains are the possible extra duality invariances of the
four-dimensional theory that are proportional to the parameters $a_A$.
These transformations have no obvious five-dimensional origin and their
presence depends entirely on the non-trivial restriction (\ref{ainv}).  We
shall denote such symmetries as {\em hidden} symmetries, to distinguish
them from those whose existence can be inferred on more general grounds, as
explained above.  However, the roots corresponding to these extra
symmetries are all located on the left half-plane in Fig.  \ref{figD}.  To
be more precise, we can decompose the symmetry algebra $\cal W$
corresponding to all the roots into eigenspaces of the generators
associated with the $\beta$ symmetry.  We then find the following
decomposition
\begin{equation}
{\cal W}= {\cal W}_{-{2/3}} +{\cal W}_0 + {\cal W}_{2/3},
\label{calW}\end{equation}
where the subscript denotes the eigenvalue with respect to the
$\beta$ symmetry. As it turns out, ${\cal W}_0$ denotes the
subalgebra associated with the parameters $\tilde B^A_{\;B}$ and
$\beta$. ${\cal W}_{{2/3}}$ contains the generators corresponding
to the parameters $b^A$ and  all possible generators
corresponding to the hidden symmetries with parameters $a_A$
belong to ${\cal W}_{-{2/3}}$. Observe that the dimension
of ${\cal W}_{-{2/3}}$ is at most equal to $n$, whereas the
dimension of ${\cal W}_{{2/3}}$ is always equal to $n$.
Unless we have maximal symmetry (i.e., unless there are $n$
independent symmetries associated with the parameters $a_A$) the
isometry group of the corresponding K\"ahler space is not
semisimple. The maximal number of isometries exists when the
curvature and the tensor $C^{ABC}$ are constant (in special
coordinates), or equivalently, when the tensor $E^G_{ABCD}$
defined in (\ref{defE}) vanishes. In that case it is known that
the corresponding K\"ahler space is symmetric \cite{BEC}.

\subsection{Special quaternionic manifolds}
\label{ss:sqm}
We now return to the general four-dimensional Lagrangian
(\ref{4dLagr}) based on a general function $F(X)$ and consider
its reduction to three space-time dimensions. In this case an
extra feature is present, because the standard (abelian) gauge field
Lagrangian in three dimensions can be converted to a scalar field
Lagrangian by means of a duality transformation. Only derivatives
of this scalar field appear, so that it has a corresponding
invariance under constant shifts. Each
four-dimensional gauge field thus gives rise to two
scalar fields with two corresponding isometries. One is its
component in the fourth dimension and the other is the scalar
field that results from the
conversion of the three-dimensional gauge field. The $n+1$
four-dimensional vector
fields $A^I_\mu$ thus give rise to $2n+2$ scalar fields, which
will be denoted by $A^I$ and $B_I$. The corresponding invariances
have parameters $\alpha^I$ and $\beta_I$.  The same conversion
can be used for
the vector field that emerges from the higher-dimensional metric,
so that the four-dimensional metric gives rise to a
three-dimensional metric and two scalar fields. These scalar
fields are denoted by $\phi$ and $\sigma$, and they also lead
to two invariances, one related to the scale transformation of
the extra coordinate with parameter $\epsilon^0$ and another one
corresponding to the converted three-dimensional vector field with
parameter $\epsilon^+$. Altogether, the
Lagrangian (\ref{4dLagr}) thus gives rise to $4(n+1)$ scalar
fields, coupled to gravity with $2n+4$ additional invariances. As
there are no vector fields anymore, the corresponding
transformations must constitute an invariance of the Lagrangian.
As this Lagrangian is still locally supersymmetric the scalar
fields must define a quaternionic non-linear sigma model
\cite{dWTNic}. The
corresponding spaces are called {\em special}
quaternionic spaces and obviously depend on a homogeneous
holomorphic function of second degree. They constitute the image
of the $\bf c$ map, as was explained in section~1. As all
quaternionic spaces, they are irreducible Einstein spaces
\cite{Ish}

The Lagrangian for this quaternionic sigma model was determined in
\cite{Sabi}, where the quaternionic structure was verified, and
in \cite{dWVP2}. It reads
\begin{eqnarray}
e^{-1}{\cal L} &=&-{\textstyle{1\over 2}} R + (zN\bar z)^{-1}
\cM_{A\overline{B}}\;\partial _\mu z^A \,\partial ^\mu\bar z^B
\nonumber \\[1mm]
&&+{\textstyle\frac{1}{4}}\phi ^{-1}(\cN +\overline{\cN})_{IJ}\,
W^I_\mu\;\overline{W}^{J\mu }\nonumber\\
&&-{\textstyle\frac{1}{4}} \phi ^{-2}\Big\{ \big(\partial _\mu
\sigma -{\textstyle\half} A^I\!\stackrel{\leftrightarrow}{\partial}
_\mu \!B_I\big)^2+\big(\partial _\mu \phi \big)^2\Big\} ,
\label{Lagr}\end{eqnarray}
where \begin{eqnarray}
W_\mu ^I&=&\big(\cN +\overline{\cN}\big)^{-1\,IJ}\left[
2\overline{\cN}_{JK}\partial _\mu A^K-i\partial _\mu B_J\right]
\end{eqnarray}
is directly related to the field strengths of the original vector
fields. Again the scalar fields are restricted to a domain $\cal
D$ in which the kinetic terms have the required signature.
{}From the fact that the field strengths $(W_\mu^I,
2i\cN_{IJ}W^J_\mu)$ transform under $Sp(2n+2,\Rbar)$ just as
$(X^I, -{1\over 2}iF_I)$, one can show that the scalar fields
$A^I$ and  $B_I$ transform linearly in the same
$(2n+2)$-dimensional representation. The
fields $\phi$ and $\sigma$ are invariant under the duality
transformations. On the other hand, as mentioned previously, the
original scalar fields $z^A$ are inert under the $2n+4$ extra
symmetries associated with the parameters $\alpha^I$, $\beta_I$,
$\epsilon^+$ and $\epsilon^0$. The full symmetry variations of
the fields will be given in \eqn{transfl}.

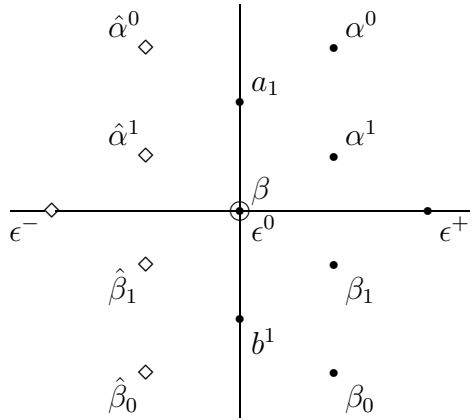
\begin{figure}[htf]
\begin{center}
\setlength{\unitlength}{0.5mm}
\begin{picture}(150,120)
\put(14,60){\line(1,0){9}}
\put(26,60){\line(1,0){110}}
\put(75,5){\line(0,1){110}}
\multiput(75,60)(50,0){2}{\circle*{2}}
\put(75,60){\circle{5}}
\put(100,16.7){\circle*{2}}
\put(100,45.6){\circle*{2}}
\put(100,74.4){\circle*{2}}
\put(100,103.3){\circle*{2}}
\multiput(75,31.1)(0,57.8){2}{\circle*{2}}
\multiput(50,16.7)(0,28.9){4}{\makebox(0,0){$\diamond$}}
\put(25,60){\makebox(0,0){$\diamond$}}
\put(22,52){\makebox(0,0)[br]{$\epsilon^-$}}
\put(78,52){\makebox(0,0)[bl]{$\epsilon ^0$}}
\put(128,52){\makebox(0,0)[bl]{$\epsilon ^+$}}
\put(103,7.7){\makebox(0,0)[bl]{$\beta _0$}}
\put(103,36.6){\makebox(0,0)[bl]{$\beta _1$}}
\put(103,77.4){\makebox(0,0)[bl]{$\alpha^1$}}
\put(103,106.3){\makebox(0,0)[bl]{$\alpha^0$}}
\put(78,22){\makebox(0,0)[bl]{$b^1$}}
\put(78,63){\makebox(0,0)[bl]{$\beta$}}
\put(78,92){\makebox(0,0)[bl]{$a_1$}}
\put(40,7.7){\makebox(0,0)[bl]{$\hat\beta _0$}}
\put(40,36.6){\makebox(0,0)[bl]{$\hat\beta _1$}}
\put(40,77.4){\makebox(0,0)[bl]{$\hat\alpha^1$}}
\put(40,106.3){\makebox(0,0)[bl]{$\hat\alpha^0$}}
\end{picture}
\end{center}
\caption{Root lattice corresponding to the isometries of an
$n=1$ special quaternionic  manifold discussed in the text.}
\label{figsolvG2}
\end{figure}

Let us again consider the root lattice corresponding to all these
symmetries. It consists of the root lattice of the duality
invariance extended by one dimension associated with the
eigenvalue of the generator $\underline\epsilon_0$ (from now on
we use the notation where generators
are denoted by their corresponding parameter, underlined with
their index raised or lowered). This leads to a root
lattice such as shown in Fig.~\ref{figsolvG2} (for the moment
ignore the generators indicated by $\diamond$), where we have exhibited the
case where the duality invariances constitute a group of rank 1 (namely
$SU(1,1)$).  The space has $n=1$ and is based on $F(X) = i(X^1)^3/X^0$.
The roots corresponding to these duality isometries are extended with the
roots belonging to the extra transformations that emerge from the reduction
to three dimensions.  Observe that there are no duality invariances
associated with the matrices $\tilde B^A_{\;B}$ in this case.  Additional
{\em hidden} symmetries are located on the left half-plane.  In the case at
hand, there are five such symmetries indicated by $\diamond$, which extend
this diagram to the root lattice of $G_{2(+2)}$.  Its solvable subalgebra
consists of the solvable subalgebra of $SU(1, 1)$, associated with the
generators $\underline\beta$ and $\underline b_1$, extended by the
generators $\underline\epsilon_0$, $\underline\epsilon_+$,
$\underline\alpha_I$ and $\underline\beta^I$ of the six extra symmetries.
In the general case we obtain a similar diagram after projecting all the
roots on a suitably chosen plane.

The algebra $\cal{V}$ corresponding to the roots of the \Ka\
space isometries and the extra symmetries, which is obviously
non-semisimple, can generally be decomposed into eigenspaces of the
generator $\underline\epsilon_0$ in the adjoint
representation.  One finds
\begin{equation}
{\cal V}= {\cal V}_0 +{\cal V}_{1/2} + {\cal V}_1 ,
\label{Qroots}
\end{equation}
where ${\cal V}_a$ denotes the eigenspace with eigenvalue $a$.
It turns out that ${\cal V}_0$ contains the generators of the duality
invariance, previously denoted by $\cal W$, supplemented by the
generator $\underline\epsilon_0$, ${\cal V}_{1/2}$
contains the $2n+2$ generators $\underline\alpha_I$ and
$\underline\beta^I$ and ${\cal V}_1$ consists of the generator
$\underline\epsilon_+$. As we shall
discuss shortly, there exist no other symmetries with
non-negative eigenvalues of $\underline\epsilon_0$, so
that the root decomposition (\ref{Qroots}) is complete in that
respect \cite{dWVP2}.

The decomposition (\ref{Qroots}) was also encountered in the
classification of {\em normal} quaternionic spaces given by \Al\
\cite{Aleks}. Normal quaternionic spaces are quaternionic spaces
that admit a transitive completely solvable group of motions. It
was conjectured in  \cite{Aleks} that the
homogeneous quaternionic spaces consist of compact symmetric
quaternionic and normal quaternionic spaces. The algebra
corresponding to the group of solvable motions in the latter case
exhibits the same decomposition as in (\ref{Qroots}), although
the spaces that emerge under the $\bf c$ map are not necessarily
normal quaternionic. According to \Al\ there are two different
types of normal quaternionic spaces characterized by their
so-called canonical quaternionic subalgebra. The first type with
subalgebra $C_1^1$ turns out to correspond to the quaternionic
projective spaces $Sp(m,1)/(Sp(m)\otimes Sp(1))$.  Their {\em
solvable} algebra ${\cal V}^{\rm s}$ decomposes as in
(\ref{Qroots}), where ${\cal V}_0^{\rm s}$ contains only the
generator $\underline\epsilon_0$, while ${\cal V}^{\rm
s}_{1/2}$ and
${\cal V}^{\rm s}_1$ have dimension $4n$ and 3, respectively. As
mentioned above, ${\cal V}_1$ is always one-dimensional, so
that the quaternionic projective spaces are {\em not} in the
image of the $\bf c$ map.

The second type has a canonical subalgebra $A_1^1$ and
the structure of the solvable algebra ${\cal V}^{\rm s}$ is as
follows:  ${\cal V}^{\rm s}_0$ is a direct sum of the generator
$\underline\epsilon_0$ and a normal
K\"ahler algebra ${\cal W}^{\rm s}$ of dimension $2n$, ${\cal
V}^{\rm s}_{1/2}$ has dimension $2n+2$ and ${\cal V}^{\rm s}_1$
has dimension 1. In order to be quaternionic, the representation
of ${\cal W}^{\rm s}$ induced by the adjoint representation of
${\cal V}^{\rm s}$ on ${\cal V}^{\rm s}_{1/2}$ must generate a
solvable subgroup of $Sp(2n+2, \Rbar)$. Hence the  special
quaternionic spaces have the same root structure (cf.
\ref{Qroots}), except that the algebra $\cal W$ is not
necessarily solvable. For a normal
K\"ahler space with duality transformations acting transitively
on the manifold, $\cal W$ has a  $2n$-dimensional solvable
subalgebra, so that the $\bf c$ map yields a normal
quaternionic space. Conversely each normal
quaternionic space with canonical subalgebra $A^1_1$
defines the basic
ingredients of a special normal K\"ahler space, encoded in its solvable
transitive group of duality transformations.
\Al's analysis thus indicates that all these spaces are in the image
of the $\bf c$ map. To establish the existence of the
corresponding four-dimensional supergravity theory, one must
prove that a holomorphic function
$F(X)$ exists that allows for these duality transformations
(i.e., that satisfies (\ref{masterdual})). This program was
carried out by Cecotti \cite{Cecotti}, who explicitly constructed
the function $F(X)$ corresponding to each of the normal quaternionic
spaces with canonical subalgebra $A^1_1$. With the exception of
the so-called minimal coupling, where $F(X)$ is a quadratic
polynomial, all functions $F(X)$ can be brought into the form
(\ref{Fd}). Recently it turned out that this conclusion also
holds for certain spaces that were missing in \Al's
classification (cf. \cite{dWVP3}).

We already alluded to possible additional (hidden)
symmetries for the special quaternionic manifolds, in analogy
with the symmetries associated with the  parameters $a_A$ in the
K\"ahler case (cf. Fig.~\ref{figD}). This question
was analyzed in \cite{dWVP2} and it was found that the hidden
symmetries are always associated to roots with eigenvalue $-1$ or
$-1/2$ with respect to the generator $\underline\epsilon_0$.  The root
lattice corresponding to all symmetries of the Lagrangian (\ref{Lagr}) thus
generally decomposes according to
\begin{equation}
{\cal V}= {\cal V}_{-1} +{\cal V}_{-1/2}
+ {\cal V}_0 +{\cal V}_{1/2} + {\cal V}_1 .\label{extQroots}
\end{equation}
As stressed previously, the roots with non-negative eigenvalues
were already completely specified in (\ref{Qroots}). The
dimensions of ${\cal V}_{-1}$ and ${\cal V}_{-1/2}$ are smaller
or equal to 1 and $2n+2$, respectively. The parameters associated
with the new
symmetries are denoted by $\epsilon^-$, corresponding to ${\cal
V}_{-1}$, and $\hat\alpha^I$ and $\hat\beta_I$, corresponding to
the generators in ${\cal V}_{-1/2}$.

The full symmetry structure of the quaternionic manifold is
encoded in a single real function $h$,
\begin{eqnarray}
h(X,\bar X,A,B) &=&-{\textstyle\frac{1}{16}} \left\{(\cB_I\bar
\cB^I)^2-{\textstyle\frac{1}{6}}\left[(
F_{IJK}\bar \cB^I\bar \cB^J\bar \cB^K)(X^L\cB_L)+
h.c.\right]\right.\nonumber\\
&&\quad\left.+{\textstyle\frac{1}{16}}(XN\bar X)
\bar \cB^I\bar \cB^JF_{IJK}N^{-1\ KL}\bar
F_{LMN}\cB^M\cB^N\right\} ,
\label{defh}
\end{eqnarray}
where
\begin{equation}
\cB_I= B_I +{\textstyle\half}{i} F_{IJ}A^J \equiv N_{IJ}\cB^J .
\label{defcB}\end{equation}
The function $h$ is a real quartic polynomial in the fields $A^I$
and $B_I$. Furthermore it is homogeneous of zeroth degree in $X$
and $\bar X$ separately, as well as invariant under duality
transformations (to see this, use the results of
appendix~\ref{app:symrepar}).

The maximal number of symmetries exists if and only if
the function $h$ is independent of $X^I$ and $\bar X^I$. In that
case the dimensions of ${\cal V}_{-1}$ and ${\cal V}_{-1/2}$ are
equal to 1 and $2n+2$, respectively, and the
isometry group is semisimple. The derivative of $h$
with respect to $X$ is proportional to a fully symmetric real
tensor $C_{IJKL}$ \cite{dWVP2},
\begin{equation}
C_{IJKL}\equiv {\textstyle\frac{1}{4}} F_{IJKL}(XN\bar X) +
(N\bar X)_{(I} F_{JKL)} -{\textstyle\frac{3}{16}} F_{M(IJ}N^{-1\,
MN}F_{KL)N}(XN\bar X) , \label{defC}
\end{equation}
which satisfies the relations
\begin{eqnarray}
X^IC_{IJKL}&=&0 ,\label{XCis0}\\
\bar X^M \frac{\partial}{\partial \bar
X^{M}} C_{IJKL}&=& -X^M \frac{\partial}{\partial X^{M}} C_{IJKL}=
 C_{IJKL} .\label{XdCC}
\end{eqnarray}
The vanishing of this tensor is a necessary and sufficient
condition for the K\"ahler manifold (as well as the associated
quaternionic manifold) to be symmetric, because of the relation
\cite{CremVP}
\begin{equation}
R^A{}_{\!BC}{}^{\!D}{}_{\!;E} =  -(z N \bar z)^{-2}\,\bar Q^{ADF}\,
{\cal C}_{FBCE}  \ ,
\label{covdR}\end{equation}
where
\begin{equation}
{\cal C}_{ABCD}
\equiv \frac{(X^0)^2}{XN\bar X}C_{ABCD} \label{defcalC}
\end{equation}
is a tensor that depends on $z$ and $\bar z$. In view of
\eqn{XCis0} ${\cal C}_{ABCD}$ contains all independent components of
$C_{IJKL}$.

When $h$ is not $X$-independent, hidden symmetries
belonging to ${\cal V}_{-1/2}$ can still exist provided that
there are linear
combinations of first-order derivatives of $h$ with respect to
$A^I$ and/or $B_I$ that are independent of $X^I$. Hence, the
situation can be summarized as follows. The hidden symmetries
correspond to the independent parameters
$\epsilon_-$, $\hat\alpha{}^I$ and $\hat\beta_I$ for which
\begin{equation}
{\cal D}h =\left( \epsilon^- +\hat\alpha{}^I\frac{\partial}{\partial
A^I} +\hat\beta_I \frac{\partial }{\partial B_I}\right) h
\label{defcalD}
\end{equation}
is independent of $X^I$ and $\bar X{}^I$.
This condition is
equivalent to the following conditions on the tensor $C_{IJKL}$,
\begin{eqnarray}
C_{IJKL}\bar \xi ^L=0  \ ,    \label{Ce1}\\
\xi ^M \frac{\partial }{\partial \bar X^M} C_{IJKL}=0  \ ,\label{Ce2}
\end{eqnarray}
where $\xi ^I$ is a function of the parameters $\hat \alpha ^I$
and $\hat \beta  _I$~:
\begin{equation}
\xi _I=\hat\beta _I+\ft12iF_{IJ}\hat\alpha^J\equiv N_{IJ}\xi ^I\ .
\label{defxi}
\end{equation}
These $2n+2$ parameters represent only
independent solutions of the above equations when the space is
symmetric and the symmetry associated with the
generator $\underline\epsilon_-$ is realized.

The conditions \eqn{Ce1} and \eqn{Ce2} can be written exclusively
in terms of ${\cal C}_{ABCD}$. Using \eqn{defcalC}
one rewrites \eqn{Ce2} as
\begin{equation}
\left( X^J\xi _J + (XN\bar X)\xi ^J \frac{\partial }{\partial \bar
X^J}\right) {\cal C}_{ABCD}=0\ .\label{Ce3}
\end{equation}
Using \eqn{XCis0} and \eqn{XdCC} and the fact that ${\cal C}_{ABCD}$
depends only on $z$ and $\bar z$, one can write
\eqn{Ce1} and \eqn{Ce3} as
\begin{eqnarray}
&&{\cal C}_{ABCD}\bar{\hat \xi }{}^{\,D} =0\ ,
\label{CeA1}\\
&&\left( z^J\xi _J+ (zN\bar z)\,\hat \xi{}^E \frac{\partial }
{\partial \bar z^E}\right) {\cal C}_{ABCD}=0\ ,\label{CeA2}
\end{eqnarray}
where $\hat \xi{}^A= \xi^A-\bar z{}^A\,\xi^0$. A more convenient
expression follows from invoking \eqn{lemma},
\begin{equation}
\hat \xi ^A = (zN\bar z)^{-1}g^{\bar AB}\xi_B - \Delta ^{-1}\left(
n^{-1}N\bar z\right) ^A \xi _I z^I   \ ,
\label{notCeA}\end{equation}
where $\Delta^{-1}=\left(N^{-1}\right)^{00}$ (alternative expressions
for $\Delta$ are given in \eqn{Delta}).
We shall return to the above conditions in section~\ref{ss:exd},
where we apply them to the manifolds
described by the functions \eqn{Fd}.

\section{The symmetry algebra of special K\"ahler and
quaternionic spaces}
\label{ss:algebra}
\setcounter{equation}{0}

In this section we discuss the algebras of the isometries of
generic special K\"ahler and quaternionic spaces and their
consequences. In
subsection~\ref{ss:skm} we have exhibited the isometries of a
generic special K\"ahler space that coincide with the generalized
duality transformations of the equations of motion corresponding
to the Lagrangian (\ref{4dLagr}). The infinitesimal
transformations act on the scalars according to \eqn{scaltra};
the matrices $C_{IJ}$, $B^I{}_J$ and $D^{IJ}$, which constitute
an infinitesimal $Sp(2n+2,\Rbar)$ transformation, are  solutions
of the consistency equation \eqn{masterdual}. We parametrize
these solutions in terms of a number of independent
transformation parameters $\omega^i$. Using the notation that the
generators corresponding to the parameters $\omega^i$ are denoted
by $\underline \omega_i$, an arbitrary infinitesimal duality
transformation is generated by
\begin{equation}
\delta(\omega) =\omega^i\,\underline\omega_i \ .
\end{equation}
As the algebra of these transformations should close, we have
\begin{equation}
[\delta (\omega_1),\delta (\omega_2)]=\delta (\omega_3)\ .
\label{deltacom}
\end{equation}
The explicit form of $\omega_3$ in terms of $\omega_1$ and $\omega_2$ can
be obtained from commuting the corresponding $Sp(2n+2,\Rbar)$ matrices
\eqn{Oinf}\footnote{Note that when the infinitesimal
transformation $\delta(\omega)$ is described by a matrix ${\cal
B}(\omega )$ (as in \eqn{Oinf}), then the commutator
corresponding to \eqn{deltacom} reads ${\cal B}(\omega _3)=\left[
{\cal B}(\omega _2),{\cal B}(\omega_1)\right]$, with $\omega_1$
and $\omega_2$ in opposite order. }.
We remind the reader that the manifold may have
additional isometries {\em not} corresponding to duality transformations.
Indeed, it was demonstrated in \cite{sssl} that
special real manifolds exist with a larger invariance group than that of
the full supergravity Lagrangian. For the special \Ka\ manifolds
we know of no examples where such a situation is realized, and for
special \Ka\ spaces based on the functions \eqn{Fd} it was
proven in \cite{sssl} that the sigma model isometries are contained in
the duality transformations. Such a proof is lacking for the
generic manifolds, but in this section we shall ignore possible
isometries of the \Ka\ space that are not contained in the group of
duality transformations.

The isometry group enlarges considerably under the \cmap, which
extends the special K\"ahler space to a special quaternionic
space. The quaternionic isometries exhibit a systematic structure
as expressed by the root decomposition \eqn{extQroots}. A
typical example of this decomposition was shown in the root
lattice Fig.~\ref{figsolvG2}. In the notation introduced in
subsection~\ref{ss:sqm}, the infinitesimal isometries for special
quaternionic manifolds are generated by
\begin{equation}
\delta =\epsilon ^0\,\underline \epsilon_0+\epsilon^+\,\underline
\epsilon_+ +\epsilon^-\,\underline \epsilon_-  +\alpha^I\,
\underline \alpha_I+\beta_I\,\underline \beta^I+\hat
\alpha{}^I\,\underline{\hat \alpha}_I+\hat\beta_I\,\underline {\hat
\beta}{}^I+\omega^i\,\underline \omega_i\ .
\label{isothree}
\end{equation}
We remind
the reader that the "extra" symmetries parametrized by
$\epsilon^+$, $\epsilon^0$, $\alpha^I$ and $\beta_I$ are realized
for any special quaternionic manifold. The generators of the
"hidden" symmetries,
belonging to the subspaces ${\cal V}_{-1}$ and ${\cal V}_{-1/2}$
of the root lattice are denoted by $\underline \epsilon_0$ and by
$\underline{\hat\alpha}{}_I$ and $\underline{\hat\beta}{}^I$,
respectively, exist whenever the appropriate conditions are
satisfied. As we discussed in subsection~\ref{ss:sqm},
the symmetry in ${\cal V}_{-1}$, parametrized by $\epsilon _-$,
is only realized for symmetric
spaces, characterized by the vanishing of the tensor $C_{IJKL}$ given in
\eqn{defC}.  If this is not the case other
"hidden" symmetries belonging to ${\cal V}_{-1/2}$ can be realized
depending on whether the conditions \eqn{Ce1} and
\eqn{Ce2} are satisfied.

Let us now turn to the infinitesimal transformations
corresponding to \eqn{isothree} and
acting on the coordinates $\phi$, $\sigma$, $A^I$, $B_I$ and
$X^I$ of the quaternionic manifold (it is more convenient to
use $X^I$ rather than $z^A$),
\begin{eqnarray}
\delta \phi &=&\phi \left(-\epsilon ^0+2\sigma \epsilon ^-
+\hat \alpha^I
B_I -\hat \beta_IA^I \right)\ ,\nonumber\\
\delta \sigma&=&\epsilon ^++\half\left(\alpha ^IB_I-\beta _IA^I\right)+
(\sigma ^2-\phi ^2)\epsilon ^-  +\sigma\left(\half
\hat \alpha^I B_I -\half\hat \beta_IA^I -\epsilon ^0\right)
+{\cal D}h\ ,\nonumber\\
\delta A^I&=&\alpha ^I+\sigma\hat \alpha ^I+ B^I_{\ J}(\omega)\,
A^J- D^{IJ}(\omega)\,B_J\nonumber\\
&&+\left(\epsilon ^-\sigma +\half \hat \alpha^JB_J
-\half\hat \beta_JA^J -\half\epsilon ^0\right)A^I -\partial
^I{\cal D}\left( h+\half \phi \cZ_2 \right)\ ,\nonumber\\
\delta B_I&=&\beta _I+\sigma\hat\beta _I+ C_{IJ}(\omega)\,A^J-
B^J_{\;I}(\omega)\,B_J \nonumber\\
&&+ \left(\epsilon ^-\sigma +\half \hat
\alpha^JB_J -\half\hat \beta_JA^J -\half\epsilon ^0\right)B_I
+\partial _I {\cal D}\left( h+\half \phi \cZ_2 \right)\ ,\nonumber\\
\delta X^I&=& B^I_{\ J}(\omega)\,X^J +{\textstyle{1\over2}}i
D^{IJ}(\omega)\, F_J\nonumber\\
&&+{\cal D} \left(-{\textstyle{1\over 2}}i \bar \cB^I\,(X^J\cB_J)
+ {\textstyle{1\over 16}}i(N^{-1})^{IJ}\cB^K\bar F_{JKL}\cB^L\,
(XN\bar X)\right) \ ,
\label{transfl}
\end{eqnarray}
where $h(X,\bar X,A,B)$ is the function given in \eqn{defh},
$\cal D$ was defined in \eqn{defcalD}, $\partial_I$ and
$\partial^I$ denote the derivatives with respect to $A^I$ and
$B_I$, respectively,  and
\begin{equation}
{\cal Z}_2\equiv \cB_I\bar \cB^I-2\frac{(X^I\cB_I)(\bar X^J\bar
\cB_J)}{XN\bar X}\ .
\end{equation}
We remind the reader that the condition for the existence of
"hidden" symmetries is that ${\cal D}h$ does not depend on  $X^I$
or $\bar X^I$.

The non-zero commutation relations of the above transformations
are as follows. First we list those that do not involve
the duality transformations,
\begin{equation}
\begin{array}{l}
[\underline{\epsilon}_0,\underline{\epsilon}_\pm ] =
\pm \underline{\epsilon}_\pm\ ,  \\[1mm]
{}[{\underline{\epsilon}}_0,\underline{\alpha}_I]=
\half \underline{\alpha}_I\ ,\\[1mm]
{}[\underline{\epsilon}_0,\underline{\beta}^I]=
\half \underline{\beta}^I\ ,\\[1mm]
{}[\underline{\epsilon}_-,\underline{\alpha}_I]=
- \underline{\hat\alpha}_I\ ,\\[1mm]
{}[\underline{\epsilon}_-,\underline{\beta}^I]
=- \underline{\hat\beta}{}^I\ ,\\[1mm]
{}[\underline{\alpha}_I,\underline{\beta}^J]
=-\delta _I^J\underline{\epsilon}_+\ ,
\end{array}
\qquad
\begin{array}{l}
[\underline{\epsilon}_-,\underline{\epsilon}_+]
=2\underline{\epsilon}_0\  ,\\[1mm]
{}[\underline{\epsilon}_0,\underline{\hat\alpha}_I]
=-\half\underline{\hat\alpha}_I\ ,\\[1mm]
{}[\underline{\epsilon}_0,\underline{\hat\beta}{}^I]
=-\half\underline{\hat\beta}{}^I\ ,\\[1mm]
{}[\underline{\epsilon}_+,\underline{\hat\alpha}_I]
=\underline{\alpha}_I\ ,\\[1mm]
{}[\underline{\epsilon}_+,\underline{\hat\beta}{}^I]
=\underline{\beta}^I\ , \\[1mm]
{}[\underline{\hat\alpha}_I,\underline{\hat\beta}{}^J]
=-\delta _I^J\underline{\epsilon}_- \ .
\end{array}
\label{comset1}
\end{equation}
Then we have
\begin{eqnarray}
[\alpha^I\,\underline{\alpha}_I+\beta _I\,\underline{\beta}^I,
\omega^i\,\underline{\omega}_i]
&=&\alpha'^I\,\underline{\alpha}_I +\beta'_I\,
\underline{\beta}^I\ , \nonumber\\{}
[\hat \alpha^I\,\underline{\hat\alpha}_I+\hat \beta _I \,
\underline{\hat\beta}{}^I,
\omega^i\,\underline{\omega}_i]&=&\hat \alpha '^I\,
\underline{\hat\alpha}_I +\hat \beta'_I\,\underline{\hat\beta}{}^I\ ,
\label{comalom}
\end{eqnarray}
where
\begin{equation}
\pmatrix{\alpha'^I\cr\noalign{\vskip3mm}  \beta'_J\cr}
= \pmatrix{B^I{}_K(\omega) & -D^{IL}(\omega)\cr \noalign{\vskip3mm}
C_{JK}(\omega) & -B^L{}_J(\omega) \cr}
\pmatrix{\alpha^K\cr\noalign{\vskip3mm}  \beta_L\cr}  \ ,
\label{comduala}
\end{equation}
and likewise for $\hat\alpha'$ and $\hat\beta'$. Finally
there are the commutators
\begin{equation}
[\alpha ^I\,\underline{\alpha}_I+\beta _I\,\underline{\beta}^I,
\hat \alpha ^I\,\underline{\hat\alpha}_I+\hat \beta _I\,
\underline{\hat\beta}{}^I]=
(\alpha ^I\hat \beta _I -\hat \alpha ^I\beta _I)\,
\underline{\epsilon}_0 +\omega^i(\alpha,\beta,
\hat\alpha,\hat\beta)\, \underline{\omega}_i\ .
\label{commalhal}\end{equation}
Explicit calculation based on \eqn{transfl} shows that the
duality transformations corresponding to the parameters
$\omega(\alpha,\beta,\hat\alpha,\hat\beta)$ correspond to the
matrices
\begin{eqnarray}
B^I_{\ J}(\alpha,\beta,\hat\alpha,\hat\beta) &=&-\half(\hat\alpha
^I\beta _J+\alpha ^I\hat\beta _J)
-\partial ^I\partial _J h''(\alpha,\beta,\hat\alpha,\hat\beta)\ ,
\nonumber\\
 C_{IJ}(\alpha,\beta,\hat\alpha,\hat\beta)&=&-\hat\beta
_{(I}\beta _{J)} +\partial _I\partial _J h''(\alpha,\beta,
\hat\alpha,\hat\beta)\ ,\nonumber\\
D^{IJ}(\alpha,\beta,\hat\alpha,\hat\beta)&=&-\hat\alpha^{(I}\alpha^{J)}
+\partial ^I\partial ^J h''(\alpha,\beta,\hat\alpha,\hat\beta) \ ,
\label{comahata}
\end{eqnarray}
where
\begin{equation}
h''(\alpha,\beta,\hat\alpha,\hat\beta)\equiv (\alpha \cdot
\partial +\beta \cdot\partial )  (\hat\alpha \cdot \partial
+\hat\beta \cdot\partial )\,h(X,\bar X,A,B) \ .
\end{equation}

{}From supersymmetry considerations, we know that the above algebra
is quaternionic. Its structure  can be visualized by a
root diagram and its Cartan subalgebra consists of the Cartan
subalgebra of the algebra corresponding to the duality
transformations (the \Ka\ algebra) and the generator
$\underline{\epsilon }_0$.  As discussed in
subsection~\ref{ss:sqm} this
leads to a typical diagram such as shown in Fig.~\ref{figsolvG2},
where the generators of the \Ka\ algebra are located on the
vertical axis. It is clear that the rank of the quaternionic
algebra is one unit higher than that of the corresponding \Ka\
algebra.

We already mentioned that when the symmetry associated with the
generator $\underline\epsilon_-$ exists, then all $2n+3$
symmetries associated with ${\cal V}_{-1}$ and ${\cal V}_{-1/2}$
are realized. When we know of the
existence of one hidden symmetry belonging to ${\cal V}_{-1/2}$,
say $\hat \alpha ^I_\star \,\underline{\hat\alpha}_I +\hat
\beta_{\star I}\, \underline{\hat\beta}{}^I$, then important
information regarding the symmetry structure follows from the
algebra. We mention four such results:
\begin{enumerate}
\item \label{it:nota0} Any other hidden symmetry belonging to
${\cal V}_{-1/2}$ and parametrized by
$\hat \alpha ^I\,\underline{\hat\alpha}_I
+\hat \beta_{I}\,\underline{\hat\beta}{}^I$ should either satisfy
$\hat \alpha\cdot\hat\beta _\star -\hat \beta \cdot
\hat\alpha_\star =0$, or $\underline{\epsilon}_-$ is also
a symmetry. The latter implies that all hidden
symmetries in ${\cal V}_{-1}$ and ${\cal V}_{-1/2}$ are realized;
in that case both the K\"ahler and the corresponding
quaternionic space are symmetric.
\item \label{it:notadn}
Also those hidden symmetries should exist that are related to
$\hat \alpha ^I_\star \,\underline{\hat\alpha}_I +\hat
\beta_{\star I}\, \underline{\hat\beta}{}^I$ by the action of the
duality transformations.
\item \label{it:h}
There exists an expression for $\big( \hat \alpha _\star
\cdot\partial +\hat \beta _\star \cdot\partial\big) h$, which
is a cubic polynomial in $A^I$ and $B_I$ with constant
coefficients that are severely restricted by the consistency
equation \eqn{masterdual}.
\item \label{it:dualex}
There exist non-trivial duality invariances.
\end{enumerate}
The first two assertions follow directly from the
last commutator in \eqn{comset1} and from \eqn{comduala}.
The last two assertions need further explanation. First consider
the commutator of
$\hat\alpha^I_\star \,\underline{\hat\alpha}_I +\hat \beta_{\star I}\,
\underline{\hat\beta}{}^I$ with $\alpha^I\,\underline{\alpha}_I
+\beta_{I}\, \underline{\beta}{}^I$, for general $\alpha$ and
$\beta$. It leads to the duality transformations parametrized by
the matrices \eqn{comahata} and implies that the third-order
derivatives of
$\left(\hat\alpha _\star \cdot\partial
+\hat\beta _\star\cdot\partial\right)h$
with respect to $A^I$ and $B_I$
must be constant (as is indeed guaranteed by the defining
condition for a hidden isometry!). Moreover the matrices
\eqn{comahata}
should satisfy the consistency condition \eqn{masterdual}, but
this fact is not so easy to exploit directly. Instead we
first explore the consequences of \eqn{comahata}
and at the end confront the result with the consistency condition.
By contracting the indices of the matrices \eqn{comahata}
with $A^I$ and $B_I$
and taking a suitable linear combination, we can use the
homogeneity of the function $h$ (which is a fourth-order
polynomial in $A$ and $B$) to obtain the expression
\begin{eqnarray}
h''(\alpha ,\beta ,\hat \alpha_\star ,\hat \beta_\star)
=-A^JB_I\,B^I{}_J(\alpha ,\beta ,\hat \alpha_\star ,\hat
\beta_\star) +\half A^IA^J \,
C_{IJ}(\alpha ,\beta ,\hat \alpha_\star ,\hat \beta_\star )
\nonumber\\
+\half B_IB_J\, D^{IJ}(\alpha ,\beta ,\hat \alpha_\star,
\hat\beta_\star)+\half (A\!\cdot\!\hat \beta_\star -B\!\cdot\!\hat
\alpha_\star) (A\!\cdot\! \beta -B\!\cdot\! \alpha).\quad
\label{H1}
\end{eqnarray}
Replacing $\alpha^I$ by $A^I$ and $\beta_I$ by $B_I$, and
using once more the homogeneity of $h$, yields
\begin{eqnarray}
(\hat \alpha_\star \cdot \partial +\hat \beta_\star \cdot
\partial )h&=&-{\textstyle{1\over3}}A^JB_I\, B^I{}_J(A,B,\hat
\alpha_\star,\hat\beta_\star ) \label{H2}\\
&& +{\textstyle{1\over6}}A^IA^J\,
C_{IJ}(A,B,\hat \alpha_\star,\hat \beta_\star)
+{\textstyle{1\over6}}B_IB_J\,
D^{IJ}(A,B,\hat \alpha_\star ,\hat \beta_\star ).
\nonumber
\end{eqnarray}
Resubstituting the above result into
\eqn{H1} shows that the matrices \eqn{comahata} can be decomposed
as follows,
\begin{eqnarray}
C_{IJ}(\alpha ,\beta,\hat \alpha_\star ,\hat \beta_\star  )&=&
C_{IJK}(\hat \alpha_\star ,\hat \beta_\star)\,
\alpha ^K + C^K{}_{IJ}(\hat \alpha_\star
,\hat \beta_\star )\,\beta _K \nonumber\\
 D^{IJ}(\alpha ,\beta ,\hat \alpha_\star ,\hat \beta_\star )&=&
D^{IJ}{}_{\!K}(\hat \alpha_\star ,\hat \beta_\star)\,
\alpha ^K + D^{IJK}(\hat \alpha_\star ,\hat \beta_\star ) \,
 \beta _K\nonumber\\
B^I{}_J(\alpha ,\beta,\hat \alpha_\star ,\hat \beta_\star )&=&-
C^I{}_{JK}(\hat \alpha_\star ,\hat \beta_\star)\,
\alpha ^K - D^{IK}{}_{\!J}(\hat \alpha_\star  ,\hat \beta_\star)\, \beta _K
   \nonumber\\
&& -\half(\hat \beta_\star \!\cdot\! \alpha
+ \beta\! \cdot\! \hat \alpha_\star) \,\delta ^I_J
-\hat\beta_{\star J}\,\alpha ^I
-\beta _J\,\hat \alpha_\star ^I
\label{exphatB},
\end{eqnarray}
where the new three-index tensors introduced on the right-hand side
are separately symmetric in upper and lower indices.
This allows us to rewrite \eqn{H2} as
\begin{eqnarray}
(\hat \alpha_\star \cdot \partial +\hat \beta_\star \cdot \partial
)h&=&{\textstyle{1\over6}}A^IA^JA^K\,
C_{IJK}(\hat \alpha_\star ,\hat \beta_\star )
+{\textstyle{1\over6}}B_IB_JB_K\, D^{IJK}(\hat
\alpha_\star ,\hat \beta_\star )\nonumber\\
&&+\half B_IB_JA^K\,
D^{IJ}{}_{\!K} (\hat \alpha_\star ,\hat \beta_\star )
+\half A^JA^KB_I \,C^I{}_{\!JK}(\hat\alpha_\star  ,\hat
\beta_\star)\nonumber\\
&& + \half (A\cdot \hat \beta_\star
+B\cdot\hat \alpha_\star  )\, (A\cdot B)\ . \label{H3}
\end{eqnarray}
The reader may wonder what we have gained here as \eqn{H3} is still
the most general expansion in terms of $A^I$ and $B_I$. However,
the matrices $B^I{}_J$, $C_{IJ}$ and $D^{IJ}$ associated with the
duality transformations are restricted by the consistency equation
\eqn{masterdual}; consequently
similar conditions exist for the tensors on the right-hand side
of \eqn{exphatB}, which, in addition, must be symmetric. To write
down these restrictions is straightforward and we refrain from
doing so. In section~\ref{ss:exd} we shall apply
this strategy to the class of spaces related to the functions
\eqn{Fd} (which do have at least one symmetry belonging to ${\cal
V}_{-1/2}$) and show how it leads to a full determination of
\eqn{H3}. Obviously, one may attempt to integrate \eqn{H3} and
in this way determine (part of) the function $h$. For the symmetric
spaces we will determine $h$ completely in this way. For
other spaces we will determine the structure of $h$ and
completely determine the terms independent of $z$, using
the duality invariance of $h$ and \eqn{H3}. In contrast,
the direct evaluation of $h$ from its definition \eqn{defh} is
difficult (see, e.g. \cite{dWVP2} where we presented
the result for the simple case corresponding to
Fig.~\ref{figsolvG2}, which required the use of a computer).

The assertion \ref{it:dualex} follows now directly from
\eqn{exphatB}, as this equation cannot be satisfied
for $ C= D= B=0$.  Therefore the corresponding K\"ahler manifold
must exhibit isometries associated with duality transformations.
\vspace{4mm}

Finally we explain under which conditions the \cmap\
preserves the homogeneity of the space.
If a manifold is homogeneous due to a particular group
of isometries, then a larger manifold in which the previous one is
embedded is also homogeneous, provided the isometries of the
submanifold can be extended to isometries of the bigger manifold
and there exist additional isometries that act transitively on the
additional coordinates.  Now consider a special \Ka\ manifold,
which is homogeneous due to the fact that certain duality
transformations act transitively on the manifold. Under the \cmap\
these transformations are extended to isometries of the quaternionic
manifold.  Moreover there are the extra symmetries corresponding
to  $\underline{\epsilon }_0$, $\underline{\epsilon }_+$,
$\underline{\alpha }_I$ and $\underline{\beta}^I$,
which by \eqn{transfl} act transitively on the additional
coordinates $\phi$, $\sigma$, $A^I$ and $B_I$.  Therefore the image of
the \Ka\ space under the \cmap\ is a homogeneous quaternionic
space.  In this respect it is
important that, after dimensional reduction, the number of new
symmetries is always larger than or equal to the number of new
coordinates.  We stress that the above arguments do not
apply to the case where the \Ka\ manifold is homogeneous owing to
isometries that are not symmetries of the full supergravity action.
So far no examples of such a manifold are known.

Conversely, consider a  manifold which is homogeneous. If there
exists a subgroup of the isometries that acts
transitively within a certain submanifold and leaves this
submanifold invariant, then the submanifold is homogeneous.
For the special quaternionic spaces we
know from \cite{dWVP2} that \eqn{transfl} comprises all the
isometries of the quaternionic space. The corresponding special
\Ka\ space, parametrized by the scalars $z^A$, is the submanifold
defined by $\phi=1$ and $\sigma = A^I=B_I=0$. The duality
transformations leave the submanifold invariant and are
isometries of the \Ka\ metric. Moreover, as the quaternionic
manifold was assumed to be homogeneous, any two points in the
submanifold are related by an isometry transformation.
Because only the action of the duality group remains inside the
submanifold, any two points of the \Ka\ manifold must be related
by a duality transformation. In other words,
the duality transformations  act transitively on the K\"ahler
manifold. It then follows that the K\"ahler manifold is also
homogeneous.

The same reasoning applies to the \rmap, where the duality
transformations are replaced by transformations that involve
vector and scalar fields and leave the $d=5$ dimensional supergravity
Lagrangian invariant.
\section{Symmetries of $d$-spaces.}
\label{ss:exd}
\setcounter{equation}{0}
This section deals with the properties of special geometries
based on the functions
\begin{equation}
F(X)= i d_{ABC} {X^AX^BX^C\over X^0}\ ,\label{Fd4}
\end{equation}
where $d_{ABC}$ are real coefficients. Some of the relevant material
was already discussed in section~\ref{ss:prel}. These functions
describe all the special real manifolds. Under the \cmap\ and the
\rmap\ they yield special \Ka\ and quaternionic spaces. The \Ka\
spaces were studied in \cite{BEC}.

After presenting some convenient
notation and other preliminaries, we recall the symmetries for
the real manifolds in subsection~\ref{ss:dRe}. Then we derive the
conditions for the  existence of hidden symmetries of the
corresponding special \Ka\ manifold (these were already quoted in
subsection~\ref{ss:skm}) and summarize the results regarding the
algebra in subsection~\ref{ss:dKa}. A large part of this section
is devoted to the symmetry structure of the corresponding
quaternionic spaces. This is done in subsection~\ref{ss:dqu},
where we give
conditions for the existence of hidden symmetries, and describe
the algebra of isometries and its main consequences. A summary of
the conditions for existence of hidden
symmetries of spaces based on \eqn{Fd4} is given in
subsection~\ref{ss:dsum}.

It is often convenient to decompose the complex fields
$z$ into real and imaginary parts,
\begin{equation}
z^A\equiv\half (y^A-ix^A)\label{zxy}\ .
\end{equation}
The domain for the variables is restricted by the
requirement that the kinetic
terms for the scalars \eqn{scalkin} and for the graviton are positive
definite. This leads to the conditions
\begin{eqnarray}
&&dxxx\equiv d_{ABC}x^Ax^Bx^C >0\nonumber\\
&&3(dxx)_A(dxx)_B-2(dx)_{AB}(dxxx)\ \mbox{ a positive definite matrix}.
\label{domaind}\end{eqnarray}
To evaluate the quantities of interest we note that
\begin{equation}
N_{00}=\textstyle{\frac{1}{2}}i(dzzz) + h.c.\ ,\qquad
N_{0A}=-\textstyle{\frac{3}{4}}i(dzz)_A +h.c. \ , \qquad
N_{AB}=\textstyle{\frac{3}{2}} (dx)_{AB} \ ,
\end{equation}
so that
\begin{equation}
N_{0A}=-\textstyle{\frac{1}{2}}N_{AB}(z+\bar z)^B  \ . \label{N0A}
\end{equation}
This implies
\begin{equation} zN\bar z=\ft 1 4 dxxx\ ,\qquad
(Nz)_A = \textstyle{\frac{3}{4}}i (dxx)_A\ ,\qquad
\left(n^{-1}Nz\right)^A =-\textstyle{\frac{1}{2}}ix^A\ ,
\end{equation}
where $\left( n^{-1}\right) ^{AB}N_{BC}=\delta^A_C$.

An important property of these models is that, within the
equivalence class of \Ka\ potentials
$K(z,\bar z)\sim K(z,\bar z)+\Lambda (z)+\bar \Lambda (\bar
z)$, there are representatives which depend only on $x$. We will
denote one such a representative by $g$
\begin{equation}
K(z,\bar z)\sim g( x) = \log dxxx \ .
\end{equation}
Therefore the metric depends only on $x$, as we found already in
\eqn{scalkin}, and is just the second derivative of $g$ with respect
to $x$. Using the notation where multiple $x$-derivatives of $g$ are
written as $g_{AB\cdots}$, the \Ka\ metric $g_{A\bar B}$
coincides with $g_{AB}$,
without the need for distinguishing holomorphic and anti-holomorphic
indices. All
multiple derivatives are homogeneous functions of $x$ and we get
the following relations
\begin{eqnarray}
&&g_A\,x^A =3\ ;\qquad g^{AB}\,g_B=-x^A\ ,\nonumber\\
&&g_A=3\,\frac{(dxx)_A}{dxxx}=-g_{AB}\,x^B\ ,\nonumber\\
&&g_{AB}=6\,\frac{(dx)_{AB}}{d\,xxx}-g_A\,g_B
=-\textstyle{\frac{1}{2}}g_{ABC}\,x^C\ ,\nonumber\\
&&g_{ABC}=6\,\frac{d_{ABC}}{dxxx}-3g_{(A}\,g_{BC)}-g_A\,g_B\,g_C
\ , \label{gderdef}
\end{eqnarray}
where $g^{AB}$ is the inverse of $g_{AB}$, which equals
\begin{equation}
g^{AB}=\for (dxxx)\left( n^{-1}\right) ^{AB}-\half
x^Ax^B\ .\label{ginvx} \end{equation}
The fourth derivative of $g$ depends on lower derivatives. This explains
why the curvature \eqn{curvC} depends only on triple derivatives of $g$.

The independent non-vanishing components of the
Christoffel connection and the Riemann tensor of a K\"ahler manifold
are generally given by
\begin{equation}
\Gamma^C_{AB}= g^{C\bar C}\, \partial_A g_{B\bar C}\ , \qquad
R^A{}_{BC\bar D}=\partial _{\bar D}\Gamma ^A_{BC}\ .
\label{Kconcur}
\end{equation}
In the case at hand the connection is purely imaginary,
\begin{equation}
\Gamma ^A_{BC}=  -\Gamma ^{\bar A} _{\bar B\bar C}=
i g^{AD}g_{BCD}\ .
\end{equation}
The (real) curvature equals
\begin{equation}
R^A{}_{BC\bar D}=R^{\bar A}{}_{\bar B\bar C D}=
\frac{\partial}{\partial x^D}\left( g^{AE}g_{EBC}\right)\ .
\end{equation}
{}From its homogeneity one proves
\begin{equation}
x^DR^A{}_{BC\bar D}=-g^{AE}g_{EBC}\ .\label{hompropR}
\end{equation}
{}From \eqn{curvC} and \eqn{defCABC} we can now verify the validity of
\eqn{covdR} for the \Ka\ spaces of this section. Using the above
expression for the connection (c.f.\eqn{Kconcur}), this leads to
an expression for the tensor $\cal C$, homogeneous in $x$,
\begin{equation}
{\cal C}_{ABCD}=
3\,d_{ABC}\,g_D
+\textstyle{\frac{9}{2}} \,g _{FD(A}\,d_{BC)G}\,g^{FG}\ ,
\end{equation}
which is not manifestly symmetric. However, it becomes fully
symmetric by using \eqn{gderdef},
\begin{equation}
{\cal C}_{ABCD}= -6d_{(ABC}\,g_{D)} +27
(g^{GH}+x^Gx^H)\,d_{H(AB}\,d_{CD)G}\,(dxxx)^{-1} .
\end{equation}
Alternative and useful representations of $\cal C$, some
not manifestly symmetric either, are easy to derive using
\eqn{gderdef},
\begin{equation}
{\cal C}_{ABCD}={\textstyle \frac{1}{18}}(dxxx)^2\,g_{AE}\,
g_{BF}\,g_{CG}\,
\frac{\partial }{\partial x^D} C^{EFG}=-6g_{DF}\,E^F_{ABCE}\,x^E
=6g_E\, E^E_{ABCD}\ .   \label{relCCE}
\end{equation}
where we used the definitions \eqn{defCABC} and \eqn{defE}. There
are many other relations between the various tensors. The
following two are particularly useful,
\begin{eqnarray}
x^A\,{\cal C}_{ABCD} &=& 0 \ , \label{Ctrans} \\
\frac{\partial }{\partial x^E} {\cal C}_{ABCD}&=&6g_{EF}\,E^F_{ABCD}\ .
\label{derCE}\end{eqnarray}

\subsection{Symmetries of the real space.}\label{ss:dRe}

The class of manifolds based on \eqn{Fd4} comprise all special real
spaces. The symmetries of the corresponding $d=5$
supergravity theory were given in \cite{GuSiTo} (see
subsection~\ref{ss:srm}). The scalar fields $h^A$, satisfying
\eqn{dhhh1}, transform as
\begin{equation}
\delta h^A= \tilde B^A_{\;B}\, h^B \ ,\label{Btild5}
\end{equation}
where the matrices $\tilde B$ define an invariance of the tensor
$d_{ABC}$ (cf. \eqn{Binv}).
However, the corresponding non-linear sigma model may possess extra
invariances, which are not symmetries of the full action.  In \cite{sssl}
we gave an example of a class of models where this is indeed the case.
Under the \cmap\ these extra transformations are not preserved,
so certain geometrical properties of the real spaces no
longer hold for the corresponding special \Ka\ spaces. In
general, only the isometries defined by \eqn{Btild5} are relevant
for the special \Ka\ and quaternionic manifolds. In \cite{sssl},
we proved that the isometries of the
corresponding \Ka\ spaces are always symmetries of the full $d=4$
supergravity theory. For the quaternionic spaces that emerge
under the \cmap\ this result is obvious as all vector fields have
then be converted to scalars.
The algebra of the transformations \eqn{Btild5} follows from the
commutators of the matrices $\tilde B$ (cf. the analogous
discussion of the algebra of the duality transformations at the
beginning of section~\ref{ss:algebra}).

\subsection{Symmetries of the \Ka\ space.}\label{ss:dKa}

The non-linear sigma models corresponding to the \Ka\ manifolds
that are in the image of the \cmap, couple to $d=4$
supergravity. The duality invariances of the corresponding
actions were studied in \cite{BEC}. As summarized in
subsection~\ref{ss:skm} these invariances lead to
isometries of the \Ka\
manifold expressed in terms of the parameters $\beta$, $b^A$,
$\tilde B^A_{\;B}$ and $a_A$, appearing in
\eqn{matrixduald}. The isometries parametrized by $\beta $ and $b^A$
\begin{equation}
\delta z^A = b^A - \textstyle{2\over 3} \beta \,z^A \, ,
\label{rztrans}
\end{equation}
exist irrespective
of the form of the symmetric tensor $d_{ABC}$ (simply because
they find their origin in the dimensional reduction from five to
four space-time dimensions (see subsection~\ref{ss:srm})).
The matrices $\tilde B$ satisfy \eqn{Binv}.
The {\em hidden} invariances associated with $a_A$ are realized for
parameters satisfying \eqn{ainv}. The derivation of the latter
result starts from the master equation \eqn{masterdual} on the
matrices \eqn{matrixduald} that parametrize the duality transformations.
This leads to
\begin{equation}
a_{(A} d_{BCD)}=-\textstyle{\frac{9}{4}} D^{EF}d_{E(AB}d_{CD)F}\
.\label{condaD} \end{equation}
Multiplying this equation with $x^Cx^D$ and with $x^Bx^Cx^D$, one
derives \eqn{DAB}, which can be re-inserted into the above
equation to yield \eqn{ainv}. However, one must ensure that the
$D^{AB}$ are constant (as the matrices \eqn{defCABC} are
constant). Perhaps somewhat surprisingly, this is again a
consequence of \eqn{ainv}. Actually, it turns out that the
following conditions are equivalent criteria for establishing the
existence of the hidden \Ka\ symmetries,
\begin{equation}
E^E_{ABCD}a_E=0 \,\Longleftrightarrow \,{\cal C}_{ABCD}\,g^{DE}\,a_E=0
\,\Longleftrightarrow\, \frac{\partial }{\partial x^D}C^{ABC}a_C=0 \ .
\label{4conda}
\end{equation}
The fact that the first condition implies the second and the
third, follows directly from \eqn{relCCE}. The second and third
condition are equivalent and imply that $a_E E^E_{ABCD}x^D=0$, again
because of \eqn{relCCE}. Because $C^{ABC}a_C$ is constant, also
$a_E E^E_{ABCD}$ is constant. Since we know that this constant
should vanish upon multiplication with $x^D$, it should itself be
zero, which is precisely the first condition \eqn{4conda}. This
proves \eqn{4conda}. Finally we remind the reader that the
conditions \eqn{4conda} are also equivalent to $R^A{}_{\!BC}{}^{\!D}a_D$
being constant. The space is symmetric if and only if the tensors
$E^E_{ABCD}$ and ${\cal C}_{ABCD}$ vanish; in that case
$R^A{}_{\!BC}{}^{\!D}$ is
thus constant (as well as covariantly constant in view of
\eqn{covdR}).\vspace{0.4cm}

The algebra of the symmetries of the \Ka\ space
was obtained in \cite{BEC}. Apart from the commutator given in the previous
subsection, the non-zero commutators are
\begin{equation}
\begin{array}{ll}
[\underline{\beta },\,\underline{b}_A]=\ft23 \underline{b}_A\ ; &
[\underline{\beta },\,\underline{a}^A]=-\ft23 \underline{a}^A\ , \nonumber\\
{}[\omega ^i\underline{\omega }{}_i,\,\underline{b}_B]=-\tilde
B^A{}_{\!B}(\omega)\,\underline{b}_A \ ; \ &
[\omega ^i\underline{\omega }{}_i,\,\underline{a}^A]=\tilde
B^A{}_{\!B}(\omega)\,\underline{a}^B\ ,\nonumber\\{}
[\underline{b}_B,\,\underline{a}^A]=\delta ^A_B\,\underline{\beta }+\omega
^{A\,i}_B\underline{\omega }{}_i \ ,
\end{array} \label{comKad}
\end{equation}
where $\omega ^{A\,i}_B$ are the transformation parameters
corresponding to
\begin{equation}
\tilde B^C{}_{\!\!D}(\omega ^A_B)=R^A{}_{\!BD}{}^C+\ft23 \delta
^C_D\,\delta ^A_B\ .
\end{equation}
These commutators clearly exhibit the  decomposition \eqn{calW}.
{}From the algebra we deduce that the existence of a symmetry
associated with parameters $a_A$ implies
the existence of duality invariances of the form
\begin{equation}
\tilde B^A{}_{\!B}(a,b)=\textstyle{\frac{4}{3}}
a_CC^{ACE}d_{BDE}\,b^D-\textstyle{\frac{1}{3}}\delta
^A_B\,b^Ca_C-b^Aa_B\ , \label{tilBa}
\end{equation}
with $b^A$ arbitrary.
This follows from taking the commutator of the hidden symmetry
with any one of the variations parametrized by $b^A$ (which are
always realized).

\subsection{Symmetries of the quaternionic space.}\label{ss:dqu}

All special quaternionic spaces possess the symmetries of \eqn{Qroots}.
The existence of possible hidden symmetries depends on existence
of solutions of \eqn{CeA1} and \eqn{CeA2}.  Here we
investigate the implications of these equations for the spaces
based on the functions \eqn{Fd4}. These quaternionic spaces are
thus in the image of the \crmap.
Using the results of the
previous subsections we first discuss a simpler version of the
equations. Subsequently, we exploit the results of
section~\ref{ss:algebra} based on the algebra of isometries,
which are very powerful. Before turning to the proofs, let us
list a number of important results. We shall show that $\hat
\beta _0$ is always associated with a symmetry, that the
conditions for $\hat \alpha ^A$ and $\hat \beta _A$ can be
written as
\begin{eqnarray}
E^E_{ABCD}\hat \alpha^D &=& 0 \label{condhatalpha}\ ,\\
E^E_{ABCD}\hat \beta _E &=& 0 \label{condhatbeta}\ ,
\end{eqnarray}
and that $\hat \alpha _0$ is only associated with a symmetry iff
the space is symmetric ($E^E_{ABCD}=0$). Below we prove
these results and exhibit how the existence of certain symmetries
implies in turn the existence of other symmetries.
Furthermore we exploit the results of section~\ref{ss:algebra} to obtain
information on the function $h$ defined
in \eqn{defh}. Some of these results were already reported in
\cite{dWVP2}.

First we note that $\hat\xi{}^A$ defined in \eqn{notCeA}
simplifies to
\begin{equation}
\hat \xi ^A= 4(dxxx)^{-1}\left( g^{AB}\xi _B+ix^A\xi _Iz^I\right) \ ,
\end{equation}
where we used relations derived at the beginning of this section
and $\Delta =-{\textstyle \frac{1}{8}}(dxxx)$. Using \eqn{Ctrans}
and the homogeneity of the tensor $\cal C$, (\ref{CeA1}) and
(\ref{CeA2}) take the form
\begin{equation}
{\cal C}_{ABCD}g^{DE}\bar \xi _E =0\ , \qquad
\xi _F g^{FE}\frac{\partial}{\partial x^E}{\cal C}_{ABCD}
=0\ .\label{Ce12}
\end{equation}
Furthermore \eqn{defxi} becomes
\begin{equation}
\xi _A =
 \hat \beta _A-3d_{ABC}z^C\hat \alpha ^B +{\textstyle \frac{3}{2}}
(dzz)_A \hat \alpha ^0\ ,  \label{xiAd}
\end{equation}
which, unlike $\cal C$, depends on both $x^A$ and $y^A$;
expansion in $y$ thus leads to a number of independent equations when
substituted into \eqn{Ce12}.
The equations for $\hat \alpha{}^A$ and $\hat \beta{}_A$ that
follow from \eqn{Ce12}, are
\begin{eqnarray}
{\cal C}_{ABCD}\,g^{DE}\,\hat \beta _E = 0 \ ;&\qquad &
\hat \beta _F\, g^{FE}\frac{\partial }{\partial x^E}
{\cal C}_{ABCD}=0\ , \nonumber\\
{\cal C}_{ABCD} \,g^{DE}\,d_{EFG}\,\hat\alpha ^F=0\ ;&\qquad &
\hat \alpha ^H \,d_{HGF}\,g^{FE} \frac{\partial }{\partial x^E}
{\cal C}_{ABCD}=0. \label{Ced4}
\end{eqnarray}
Furthermore, we can only have a
symmetry associated with the parameter $\hat\alpha{}^0$ provided
that ${\cal C}=0$.
This implies that the corresponding \Ka\ space is
symmetric (and, as it turns out from explicit application of the
\cmap, so is the corresponding quaternionic space).

None of the above equations depend on
$\hat\beta_0$, so that the hidden symmetry corresponding to this
parameter is {\em always} realized. Subsequently consider the
equations \eqn{Ced4} proportional to $\hat\beta$. It turns out
that the first equation coincides with the second
condition for the hidden \Ka\ symmetry characterized by the
parameter $a_A$ in \eqn{4conda}, while according to \eqn{derCE}
and \eqn{4conda}, the second equation is just equivalent to the
first. This establishes that the condition for the existence of
hidden quaternionic symmetries associated with the parameters
$\hat\beta _A$  coincides with the condition for
the hidden \Ka\ symmetries parametrized by $a_A$, which, again
according to \eqn{4conda}, is precisely \eqn{condhatbeta}.
Consequently, every hidden \Ka\ symmetry implies the existence of
a hidden quaternionic symmetry associated with parameters
$\hat\beta_A$.

Following the same arguments, the equations \eqn{Ced4} for the
symmetries associated with the parameters $\hat \alpha ^A$ are
that the parameters
$\hat \beta _A=d_{ABC}\hat \alpha ^B$ satisfy \eqn{condhatbeta} and
thus constitute hidden symmetries for all index values for $C$.
According to \eqn{4conda} this means that $E^E_{ABCD}\,d_{EFG} \,
\hat\alpha{}^F=0$. The latter proves that parameters $\hat
\alpha $ satisfying \eqn{condhatalpha} are symmetries. Indeed,
the identity \begin{equation}
E^E_{ABCD}\,d_{FGE} +2 E^E_{F(ABC}\,d_{D)GE} -3E^E_{FG(AB}\,
d_{CD)E} =0 \ ,
\end{equation}
multiplied with $\hat\alpha ^F$, shows that for parameters satisfying
\eqn{condhatalpha}, $\hat \beta _A=d_{ABC}\hat \alpha ^B$ are symmetries
for all index values of $C$. This proves that \eqn{condhatalpha} is a
sufficient condition. Its necessity will be derived shortly.

Many of the above results and their implications can be
obtained from the algebra of isometries discussed in
section~\ref{ss:algebra}. The fact that the $\hat \alpha ^0$
symmetry does not
exist unless the space is symmetric, can be understood from the
result~\ref{it:nota0}  and the existence of the $\hat\beta_0$
symmetry. For symmetric spaces all hidden symmetries are
realized; therefore we will not discuss them further and ignore
the $\hat \alpha ^0$ symmetry. For non-symmetric spaces the
independent parameters $\hat \alpha{}^A$ and $\hat\beta_A$ are
subject to the constraint $\hat \alpha{}^A\,\hat \beta_A=0$.
The result~\ref{it:notadn} now takes the following form. From
\eqn{comalom} and \eqn{comduala} we conclude that if we know of
the existence of symmetries with parameters $\hat \alpha_\star^I,
\,\hat \beta_{\star I}$, then hidden symmetries must be realized
with the following parameters $\hat\alpha ^I,\, \hat \beta _I$,
\begin{equation}
\pmatrix{\hat\alpha ^0 \cr\hat\alpha ^A\cr\hat\beta _0\cr\hat\beta_A\cr}=
\pmatrix{\beta &a_B&0 & 0 \cr
        b^A&\tilde B^A{}_B+\ft13\beta \delta ^A_B & 0 &\ft49C^{ABC}a_C\cr
         0 &       0                         & -\beta &-b^B\cr
     0 & 3d_{ABC}b^C  & -a_A & -\tilde B^B{}_A-\ft13 \beta \delta^A_B\cr}
\pmatrix{\hat\alpha_\star^0\cr\hat\alpha_\star^A\cr
\hat\beta_{\star 0}\cr\hat \beta _{\star A}\cr}\ ,
\label{bigmatd}
\end{equation}
where the parameters $\beta$, $b^A$, $\tilde B{}^A_{\;B}$ and
$a_A$ parametrize duality transformations. We list three
consequences of this formula:
\begin{enumerate}\label{conclald}
\item We established that the symmetry associated with the
parameter  $\hat \beta_{\star 0}$ is always realized.  Therefore,
if there are duality transformations characterized by parameters
$a_A$, there must be extra symmetries with $\hat
\beta_A\propto a_A$. This explains why the conditions for the
existence of $\hat \beta _A$ symmetries are the same as those for
$a_A$ symmetries, as derived above.
\item If $\hat \alpha ^A_\star$ corresponds to an isometry, so
does $\tilde B{}^A_{\;B}\,
\hat \alpha ^B_\star$.  By using the duality transformations
proportional to $b^A$ (which are always realized) we conclude
that the parameters $\hat\beta_A\propto d_{ABC}\hat \alpha
^B_\star$ should correspond to an isometry as well, in accord
with the result found above.
\item If there are symmetries corresponding to the parameters
$\hat \beta _{\star B}$ then there should also be symmetries
corresponding to the parameters $\tilde B^B{}_{\!A}\,\hat \beta
_{\star B}$. Furthermore we can make use of the fact that for
every duality transformations proportional to $a_A$, there exist a
hidden quaternionic symmetry associated with the parameters $\hat
\beta _{\star A}\propto a_A$. Then there must be an isometry
associated with the parameters
\begin{equation}
\hat \alpha ^A=C^{ABC}\,\hat\beta_{\star B}\hat \beta_{\star C}
\label{alphabebe}
\end{equation}
for all independent parameters $\hat \beta _{\star A}$ associated
with a hidden quaternionic symmetry.
\end{enumerate}

We now explore the commutator \eqn{commalhal} discussed in
section~\ref{ss:algebra}. The resulting duality transformation,
parametrized by \eqn{exphatB}, should be in accord with the
general form of the duality transformations of the \Ka\ spaces
based on \eqn{Fd4}, specified in \eqn{matrixduald} and \eqn{DAB},
for arbitrary $\alpha $ and $\beta $. This requirement leads to
surprisingly restrictive conditions for the
coefficients in \eqn{exphatB}. The non-vanishing coefficients are
given by
\begin{equation}
\begin{array}{l}
C_{ABC}(\hat \alpha ,\hat\beta )=-3d_{ABC}\hat \beta _0\ ,\\
C^C{}_{AB}(\hat \alpha ,
\hat\beta)=-\textstyle{\frac{4}{3}}d_{ABD}C^{CDE}\hat\beta _E\ , \\
C^0{}_{AB}(\hat \alpha ,\hat\beta )=-3d_{ABC}\hat\alpha ^C\ ,
\end{array}  \quad
\begin{array}{l}
D^{ABC}(\hat \alpha ,\hat\beta )=\textstyle{\frac{4}{9}}
C^{ABC}\hat \alpha ^0 \ ,\\
D^{AB}{}_C(\hat \alpha ,\hat\beta )= -\textstyle{\frac{4}{3}}
C^{ABD}d_{CDE}\hat\alpha^E\ ,\\
D^{AB}{}_0(\hat \alpha ,\hat\beta )=\textstyle{\frac{4}{9}}
C^{ABC}\hat\beta _C \ ,
\end{array}
\label{CD3d}
\end{equation}
while the parameters of the duality
transformations in the commutator \eqn{commalhal} are equal to
\begin{eqnarray}
\beta(\alpha ,\beta ,\hat \alpha ,\hat\beta )&=&-\half \hat \beta_I
\alpha^I-\hat \beta _0\alpha ^0 -\half \beta_I\hat
\alpha^I- \beta _0\hat\alpha ^0\nonumber\\
b^A(\alpha ,\beta ,\hat \alpha ,\hat\beta )&=&-\textstyle{\frac{4}{9}}
C^{ABC}\beta
_B\hat\beta _C-\hat\beta _0\alpha ^A-\beta _0\hat\alpha ^A\nonumber\\
a_A(\alpha ,\beta ,\hat \alpha ,\hat\beta )&=&3d_{ABC}\alpha ^B\hat\alpha
^C-\hat\alpha ^0\beta _A-\alpha^0\hat\beta _A\nonumber\\
\tilde B^A{}_B(\alpha ,\beta ,\hat \alpha ,\hat\beta )&=&
\textstyle{\frac{4}{3}}C^{ACD}d_{BCE}(\hat \beta _D\alpha ^E
+\beta _D\hat\alpha ^E )\nonumber\\ &&
-\textstyle{\frac{1}{3}}\delta ^A_B (\hat \beta _C\alpha
^C+\beta _C\hat \alpha ^C)-\hat \beta _B\alpha ^A-\beta _B\hat\alpha ^A .
\label{pardualcom} \end{eqnarray}
The above parameters should satisfy
\eqn{Binv} and \eqn{ainv} for all $\alpha ^I$ and $\beta _I$. (Observe
that $\hat \beta _0$ does not appear in the expressions
for $a_A$ and $\tilde B^A{}_B$, so it is left unrestricted as it
should.) This gives rise to the equations
\begin{equation}
\hat\alpha{}^0 E^E_{ABCD} =E^E_{ABCD}\, d_{EFG}\,\hat\alpha{}^F =
E^E_{ABCD} \,\hat\beta_E= E^E_{ABCD}\,\hat\alpha{}^D =0\ .
\label{condalbet}
\end{equation}
This confirms the previous results and finally establishes the
necessity for the condition \eqn{condhatalpha}.

\vspace{0.5cm}

Finally we exploit the result 3 on page~\pageref{it:h} to further
determine the function $h$. Using \eqn{CD3d} we may write \eqn{H3}
as
\begin{eqnarray}
(\hat \alpha{}^I \partial_I+\hat \beta_I\partial^I) h&=&
\hat\beta_0\left\{-\ft12 d_{ABC}\,  A^AA^BA^C+ \ft12 A^0 A^I
B_I\right\}\label{hderivative}\\
&&+\hat\alpha^0\left\{\ft2{27} C^{ABC}\,B_AB_BB_C +\ft12
A^IB_IB_0\right\}\nonumber\\
&&+\hat\beta_A \left\{
\ft{2}{9}A^0\,C^{ABC}\,B_BB_C\,
-\ft{2}{3}A^DA^E\,d_{BDE}\,C^{ABC} \,B_C
+\half A^AA^IB_I \right\}\nonumber\\
&&+ \hat\alpha{}^A\left\{-\ft23 A^C\,d_{ABC}\,C^{BDE}\,B_DB_E
-\ft23 A^BA^C\,d_{ABC}\,B_0 +\ft12 A^IB_IB_A \right\}\,,
\nonumber
\end{eqnarray}
for the independent parameters $\hat\alpha$
and $\hat\beta$ that are associated with a hidden symmetry.
We recall that the right-hand side of this equation does not
depend on $z^A$ and $\bar z{}^A$. In the case at hand, we know that
the symmetry associated with $\hat \beta _0$ is always
realized, so that the derivative of $h$ with respect to $B_0$
equals
\begin{equation}
\partial ^0 h = -\half d_{ABC} A^AA^BA^C+\half A^0 A^I B_I\ .
\label{B0der}
\end{equation}
This determines the function $h$ up to a quartic polynomial in
$A^0$, $A^A$ and $B_A$, with $z$- and $\bar z$-dependent
coefficient functions. However, $h$ should be
invariant under duality transformations. In this case we know
that the transformations parametrized by $\beta$ and $b^A$ are
always realized. For the fields $A^0,A^A, B_0, B_A$, these
transformations can be read off immediately from
\eqn{bigmatd}, while for $z$ and $\bar z$ they are given in
\eqn{rztrans}. To further determine the function, such that
the result is manifestly invariant under these duality
transformations, we express it in terms of fields that are
invariant under the duality transformations proportional to
$b^A$, namely $x^A$, $A^0$ and
\begin{eqnarray}
\tilde A{}^A&=& A^A-\ft12 A^0 \,y^A\ ,\nonumber\\
\tilde B_A&=& B_A-\ft32 (d\,y)_{AB}\,A^B + \ft38 (d\,yy)_A\, A^0
\ ,\\
\tilde B_0&=& B_0 +\ft12 y^A\,B_A -\ft38(d\,yy)_A\,A^A +\ft1{16}
(d\,yyy)\,A^0\ ,\nonumber
\end{eqnarray}
where $y^A$ and $x^A$ are proportional to the real and imaginary
part of $z^A$ (cf. \eqn{zxy}). Expressing a function in terms of
these variables and $y^A$, the invariance
of $h$ under the $b^A$ transformations implies that there is no
explicit dependence on $y^A$, so we may write $h(A^0,\tilde A^A,\tilde
B_0,\tilde B_A,x^A)$. Because
$\frac{\partial }{\partial B^0}h= \frac{\partial }{\partial\tilde  B^0}h$,
and the right-hand side of \eqn{B0der} does not change under the
replacement of $A^A$, $B_A$ and $B_0$ by $\tilde A{}^A$, $\tilde
B_A$ and $\tilde B_0$ (due to the fact that this expression
is invariant under the duality transformations proportional to
$b^A$), we have thus determined the dependence of $h$ on $\tilde
B^0$. The other terms in $h$ are quartic in $A^0$,
$\tilde A^A$ and $\tilde B_A$, and there are 15 combinations of these.
Under duality transformations parametrized by the parameter
$\beta$ these fields scale with weight 1, $\ft13$ and $-\ft13$,
respectively, while $x^A$  carries weight
$-\ft23$. Each coefficient function of $x$ is thus homogeneous of a
suitable degree. There appear functions of all degrees from $-2$ to 6.
For instance, the contribution to $h$ that is of zeroth degree in
$x$, depends on two such functions, $h^{ABC}(x)$ and
$h^{AB}_{CD}(x)$,
\begin{eqnarray}
h^{(0)}&=&-\half  d_{ABC}\,\tilde A{}^A\tilde A{}^B\tilde A{}^C\,
\tilde B_0 +\half
A^0\,\tilde A{}^A\,\tilde B_A \,\tilde B_0
+\textstyle{\frac{1}{4}}(A^0)^2(\tilde B_0)^2\nonumber\\
&& +A^0\,\tilde B_A\,\tilde B_B\,\tilde B_C \,h^{ABC}(x)
+\tilde A{}^A\,\tilde A{}^B\,\tilde B_C\,\tilde B_D\,
h^{CD}_{AB}(x) \ .
\end{eqnarray}
This function is manifestly invariant under duality
transformations associated with the parameters $b^A$ and $\beta$
and satisfies \eqn{B0der}. The remaining functions are $\tilde
B_0$ independent and of non-zero degree in $x$.

When there are additional symmetries known proportional to
parameters $\hat\beta_A$ and/or $\hat\alpha{}^I$ one can again
follow the same strategy and use \eqn{hderivative} to  further
restrict the function $h$. A special case is that of the
symmetric space, where the maximal number of symmetries is
realized, so that \eqn{hderivative} holds for all
$\hat\alpha{}^I$ and $\hat\beta_I$ separately. In that case all
first-order derivatives with respect to $A_I$ and $B_I$ are known
and we can integrate the result to find the function $h$,
\begin{eqnarray}
h&=&-\half d_{ABC}\,A^A\,A^B\,A^C \,B_0 +\half
A^0\,A^A\,B_A \,B_0 +\textstyle{\frac{1}{4}}(A^0)^2(B_0)^2\label{hsym}\\
&& +\textstyle{\frac{2}{27}}A^0 \,B_A \,B_B \,B_C \,
C^{ABC}-\textstyle{\frac{1}{3}}
A^A\,A^B\,B_C\,B_D\,d_{ABE}C^{ECD}+\textstyle{\frac{1}{4}}A^A\,
B_A\,A^B\,B_B.
\nonumber
\end{eqnarray}
One can check that this result is invariant under all duality
transformations. In particular it is invariant under the duality
transformations proportional to the parameters $b^A$. This
implies that the replacement of $A^A$, $B_A$ and $B_0$ by $\tilde
A{}^A$, $\tilde B_A$
and $\tilde B_0$ leaves the function unaffected. To show this
it is important that the tensor $E^E_{ABCD}$ vanishes.
For $n=1$ with $d_{111}=C^{111}=1$ the above formula for $h$ confirms
the result of a computer calculation presented in \cite{dWVP2}.

\subsection{Summary}\label{ss:dsum}
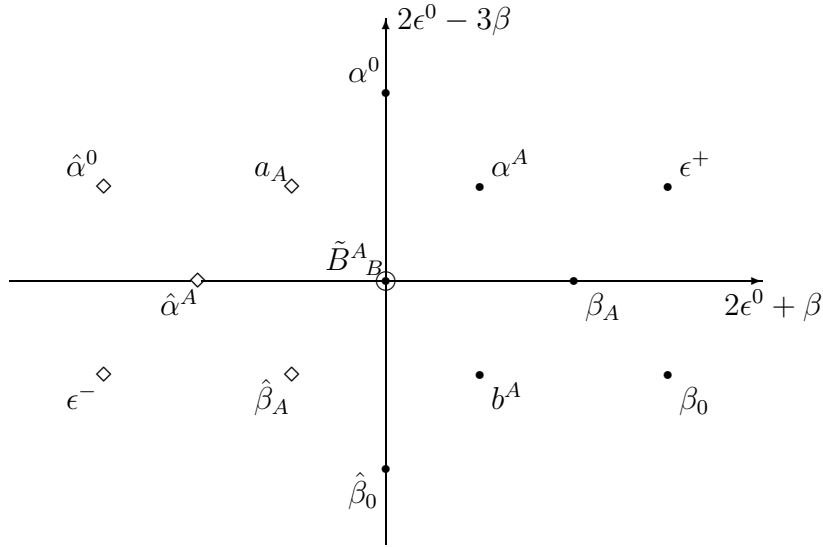
\begin{figure}[htf]\caption{Roots of
quaternionic $d$-spaces  in the
$\beta$-$\epsilon^0$ plane} \label{rootspec}\begin{center}
\setlength{\unitlength}{0.5mm}
\begin{picture}(200,140)(-100,-70)
\put(-100,0){\line(1,0){48}}
\put(-49,0){\vector(1,0){149}}
\put(0,-70){\vector(0,1){140}}
\multiput(0,-50)(0,50){3}{\circle*{2}}
\put(0,0){\circle{5}}
\multiput(-75,-25)(50,0){2}{\makebox(0,0){$\diamond$}}
\multiput(-75,25)(50,0){2}{\makebox(0,0){$\diamond$}}
\put(-50,0){\makebox(0,0){$\diamond$}}
\multiput(25,-25)(50,0){2}{\circle*{2}}
\multiput(25,25)(50,0){2}{\circle*{2}}
\put(50,0){\circle*{2}}
\put(90,-9){\makebox(0,0)[bl]{$2\epsilon ^0+\beta $}}
\put(3,67){\makebox(0,0)[bl]{$2\epsilon ^0-3\beta $}}
\put(-85,-34){\makebox(0,0)[bl]{$\epsilon ^-$}}
\put(-16.1,3){\makebox(0,0)[bl]{$\tilde B^A{}_{\!B}$}}
\put(-10,-59){\makebox(0,0)[bl]{$\hat\beta _0$}}
\put(-35,-34){\makebox(0,0)[bl]{$\hat\beta _A$}}
\put(-60,-9){\makebox(0,0)[bl]{$\hat\alpha^A$}}
\put(-85,28){\makebox(0,0)[bl]{$\hat\alpha^0$}}
\put(78,28){\makebox(0,0)[bl]{$\epsilon ^+$}}
\put(78,-34){\makebox(0,0)[bl]{$\beta _0$}}
\put(53,-9){\makebox(0,0)[bl]{$\beta _A$}}
\put(28,28){\makebox(0,0)[bl]{$\alpha^A$}}
\put(-10,53){\makebox(0,0)[bl]{$\alpha^0$}}
\put(28,-34){\makebox(0,0)[bl]{$b^A$}}
\put(-35,28){\makebox(0,0)[bl]{$a_A$}}
\end{picture}
\end{center}
\end{figure}

For the $d$-spaces the symmetries that are realized for the
various types of special manifolds are summarized in Fig.~\ref{rootspec}
and in Table~\ref{symd}.
The figure can be compared to Fig.~\ref{figsolvG2}, which
corresponds to the case where the isometry algebra of the
quaternionic space equals $G_2$, except that we have changed
the axes. The corresponding quaternionic
space has rank 2, and is a $d$-space, i.e., it is in the image of
the \crmap.

\begin{table}[tf]
\caption{Symmetries of $d$-spaces}
\label{symd}\begin{center}\begin{tabular}{||ccc|rr|rr||}
\hline
real space&K\"ahler space  & quaternionic space
&$\underline{\beta}$ &$\underline{\epsilon}{}_0 $&
$2\underline{\epsilon}{}_0+ \underline{\beta}$ &
$2\underline{\epsilon}{}_0-3 \underline{\beta} $ \\[2mm] \hline
               &           & $\epsilon ^+$&$0$&$1$&2 &2 \\[2mm]
               &           & $\alpha ^0$&$-1$ & $\half$ &0 &4 \\[2mm]
               &           & $\beta _0$&$1$ & $\half$ & 2&$-2$ \\[2mm]
               &           & $\alpha ^A$&$-\frac{1}{3}$&$\half$
                                                    &$\ft23$&2\\[2mm]
               &           & $\beta _A$&$\frac{1}{3}$ & $\half$
                                                    &$\ft43$&0\\[2mm]
               & $b^A   $  & $b^A$&$\frac{2}{3}$ & $0$
                                                   &$\ft23$&$-2$ \\[2mm]
               & $\beta $  & $\beta,\ \epsilon ^0 $& $0$ & $0$ &0&0\\[2mm]
               &           & $\hat \beta _0$&$1$ &$-\half$&0&$-4$ \\[2mm]
               \hline
$\tilde B^A{}_{\!B}$& $\tilde B^A{}_{\!B}$ & $\tilde
B^A{}_{\!B}$&$0$ & $0$ & 0&0  \\[2mm] \hline
               & $a_A$     & $a_A$    &$-\frac{2}{3}$ & $0$
                                                  &$-\ft23$&2 \\[2mm]
               &           & $\hat \beta_A$   &$\frac{1}{3}$ & $-\half$
                                                  &$-\ft23$&$-2$ \\[2mm]
\hline &           & $\hat \alpha ^A$ &$-\frac{1}{3}$ & $-\half$
                                                  &$-\ft43$&0 \\[2mm]
\hline &           & $\hat \alpha ^0$&$-1$ & $-\half$ &$-2$ &2 \\[2mm]
               &           & $\epsilon ^-$&$0$ & $-1$ &$-2$ &$-2$\\[2mm]
 \hline
\end{tabular}\end{center}
\end{table}
The symmetries with $2\epsilon^0+\beta\geq 0$ that are listed above the first
horizonal line in the table, are realized for arbitrary coefficients
$d_{ABC}$. All those roots are located on the right half-plane
in the figure.

The symmetries associated with $\tilde B^A_{\;B}$, which are the
remaining symmetries with $2\epsilon^0+\beta=0$, are listed
in the table between the next two horizontal lines. The existence
of these symmetries is related to solutions of the equation
\begin{equation}
d_{D(AB}\tilde B^D{}_{C)}=0\ .
\end{equation}
The latter represent all invariances of the full $d=5$
supergravity theory that contains a non-linear sigma model with
the real space as a target space. It was shown in
\cite{sssl} that the real manifold can have additional
isometries, which are not preserved by the supergravity
interactions and
therefore do not reappear as isometries of the corresponding \Ka\
and quaternionic manifolds. Therefore those symmetries are not
listed here.

The Cartan subalgebra of the $\tilde B^A{}_B$ transformations
gives rise to additional dimensions of the root diagram, which are
not depicted in Fig.~\ref{rootspec}. Their presence depends on
the particular choice for the $d$-coefficients. In the next
section, these generators will be
specified for the homogeneous $d$-spaces. The Cartan subalgebra
of the corresponding \Ka\ space consists of the Cartan
subalgebra of the real space complemented by the generator
$\underline{\beta}$. The Cartan subalgebra of the quaternionic
space consists of the \Ka\ Cartan subalgebra extended with the
generator $\underline{\epsilon}{}^0$.
Hence the rank of the isometry algebra always increases by one unit.

The symmetries with $2\epsilon ^0+\beta =-\ft23$ are
related to the solutions of
\begin{equation}
E^E_{ABCD}\,a_E= E^E_{ABCD}\hat\beta_E=0\ .
\end{equation}
Therefore, the condition determines the existence of hidden
symmetries for both the \Ka\ and the quaternionic manifold. In
Fig.~\ref{rootspec} one can see that the $\hat\beta{}_0$ symmetry,
which is always realized, relates the  $a_A$  to the
$\hat\beta{}_A$ symmetries.

The symmetries with $2\epsilon ^0+\beta =-\ft43$ are related to
solutions of
\begin{equation}
E^E_{ABCD}\hat \alpha ^D=0\ . \label{condhatal}
\end{equation}
Finally, the symmetries with $2\epsilon ^0+\beta =-2$ exist
only for the symmetric spaces, which means that
\begin{equation}
E^E_{ABCD}=0\ .
\end{equation}

The existence of symmetries with a certain (negative) eigenvalue
of $2\underline{\epsilon}_0+\underline{\beta}$ implies always
that symmetries exist with an eigenvalue that is $\ft23$ larger.
For instance, the
existence of $a_A$ symmetries implies the existence of a $\tilde
B$ symmetry of the form \eqn{tilBa} for arbitrary $b^A$, and
presence of $\hat\alpha ^A$ symmetries implies that
$\hat\beta_{A}=d_{ABC}\hat\alpha ^B$ is a symmetry for all $C$.

\section{Homogeneous $d$-spaces. }
\setcounter{equation}{0}\label{ss:gamma}

For a homogeneous manifold the isometry group acts transitively,
which means that every two points of the manifold are related by
an isometry transformation. Therefore the manifold itself is
(locally) determined by the isometry group. In this section we
want to study the isometry group $G$
for homogeneous spaces based on the coefficients $d_{ABC}$. Again these
spaces come in three varieties, namely real, \Ka\ and
quaternionic. In \cite{sssl} they were called `very special'.
The homogeneous quaternionic spaces consist of the (symmetric)
quaternionic projective spaces, the symmetric spaces that are
related to the projective complex spaces via the \cmap\ and a
subset of the $d$-spaces. The corresponding \Ka\ and real
spaces are homogeneous as well, according to the arguments
presented at the end of section~\ref{ss:algebra}. For the \Ka\
spaces we know from \cite{sssl} that there are no other
homogeneous spaces based on $d_{ABC}$ coefficients, while all
symmetric special K\"ahler spaces are contained in this class
according to \cite{CremVP}. Whether or not there are real special
spaces that are homogeneous and do not lead to homogeneous \Ka\
and quaternionic spaces via the $\bf r$ and the \cmap, is not
known. For such spaces the transitive symmetry (sub)group cannot
be extended to a symmetry of the corresponding supergravity
action; at present no example of such a space is known.\footnote{In
\cite{sssl} examples were discussed of real special spaces whose
isometries were not all contained in the invariance group of the
supergravity action; in spite of that the latter group still acted
transitively on the space.} In this section we discuss all
homogeneous quaternionic spaces based on $d_{ABC}$ coefficients and
their real and \Ka\ partners.

The $d_{ABC}$ coefficients that give rise to homogeneous spaces were
classified in \cite{dWVP3}.  The solutions are denoted by $L(q,
P)$, where $q$ and $P$ are integers with $q\geq -1$ and $P\geq
0$.  For $q$ equal to a multiple of 4,
there exist additional solutions denoted by $L(4m,P,\dot P)$,
with $m\geq 0$ and $P,\dot P\geq 1$.  Here $L(4m,P,\dot P)=L(4m,
\dot P,P)$.  Some of the corresponding spaces were assigned
various names in the literature \cite{Aleks,Cecotti,dWVP3}, which
are given in in Table~\ref{homsp}, where the spaces discovered in
\cite{dWVP3} are indicated by a $\star$.
\begin{table}[htb]
\begin{center}
\begin{tabular}{|l|ccc|c|}\hline
$C(h)$&real & K\"ahler & quaternionic   \\
\hline&&&\\[-3mm]
$L(-1,0)$&$SO(1,1)$&$\left[\frac{SU(1,1)}{U(1)}\right]^2$&$\frac{SO(3,4)}{(
S U ( 2 ) ) ^ 3 } $ \\[2mm]
$L(-1,P)$&$\frac{SO(P+1,1)}{SO(P+1)}$& $\star$ & $\star$ \\[2mm]
\hline&&&\\[-3mm]
$L(0,0)$&$[SO(1,1)]^2$&$\left[ \frac{SU(1,1)}{U(1)}\right] ^3$&$
\frac{SO(4,4)}{SO(4)\otimes SO(4)} $\\[2mm]
$L(0,P)$&$\frac{SO(P+1,1)}{SO(P+1)}\otimes
SO(1,1)$&$\frac{SU(1,1)}{U(1)}\otimes
\frac{SO(P+2,2)}{SO(P+2)\otimes SO(2)}$&$
\frac{SO(P+4,4)}{SO(P+4)\otimes SO(4)} $\\[2mm]
$L(0,P,\dot P)$&$Y(P,\dot P)$&$K(P,\dot P)$&$W(P,\dot P)$ \\[2mm]
$L(q,P)$&$X(P,q)$&$H(P,q)$&$V(P,q)$\\[2mm]
$L(4m,P,\dot P)$& $\star$ & $\star$& $\star$\\[2mm]
$L(1,1)$&
$\frac{S\ell(3,\Rbar)}{SO(3)}$&$\frac{Sp(6)}{U(3)
}$&$\frac{F_4}{USp(6)\otimes SU(2)}$\\[2mm]
$L(2,1)$&
$\frac{S\ell(3,\Cbar)}{SU(3)}$&$\frac{SU(3,3)}{SU(3)\otimes
SU(3)\otimes U(1)}$&$\frac{E_6}{SU(6)\otimes SU(2)}$\\[2mm]
$L(4,1)$&
$\frac{SU^*(6)}{Sp(3)}$&$\frac{SO^*(12)}{SU(6)\otimes
U(1)}$&$\frac{E_7}{\overline{SO(12)}\otimes SU(2)}$\\[2mm]
$L(8,1)$&
$\frac{E_6}{F_4}$&$\frac{E_7}{E_6\otimes
 U(1)}$&$\frac{E_8}{E_7\otimes SU(2)}$\\[2mm]
\hline \end{tabular}
\end{center}
\caption{Homogeneous special real spaces and their corresponding
K\"ahler and quaternionic spaces. The rank of the real spaces is
equal to 1 (above the line) or 2 (below the line). The rank of
the corresponding \Ka\ and quaternionic manifolds is increased by
one or two units, respectively. The integers $P$, $\dot P$, $q$
and $m$ can take all values $\geq 1$. } \label{homsp}
\end{table}
The symmetric spaces are the three varieties corresponding to
$L(-1,0)$, $L(0,P)$, $L(1,1)$, $L(2,1)$, $L(4,1)$, and $L(8,1)$,
and the real spaces corresponding to $L(-1,P)$.  For those cases
the isometry group $G$ is given. For the non-symmetric spaces the
isometry group $G$ is not semisimple; the isotropy group $H$ is
always its maximal compact subgroup.  The aim of this section
is to clarify the structure of these isometry groups $G$.

The $d_{ABC}$ coefficients for these spaces can be specified as
follows (we use here the parametrization chosen in the last
section of \cite{dWVP3}, which is different from the so-called
`canonical parametrization'). First we decompose the
coordinates $h^A$ into
$h^1$, $h^2$, $h^\mu$ and $h^i$, where the indices
$\mu$ and $i$ run over $q+1$ and $r$ values, respectively. Hence we have
$n= 3+q+r$.
The non-zero components of $d_{ABC}$ are
\begin{equation}
d_{122}=1\ ;\qquad d_{1\mu \nu }=-\delta _{\mu \nu }\ ;\qquad
d_{2ij}=-\delta _{ij}\ ;\qquad d_{\mu ij}=\gamma _{\mu ij}\ ,
\end{equation}
so that $C(h)$ can be written as
\begin{equation}
C(h) = 3\Big\{ h^1\,
\big(h^2\big)^2 -h^1\,\big(h^\mu\big)^2 -h^2\,\big(h^i\big)^2
+\gamma_{\mu ij}\,h^\mu\, h^i\,h^j\Big\} \ .\label{genC1}
\end{equation}
The coefficients $\gamma_{\mu ij}$ are the generators of a
$(q\!+\!1)$-dimensional real Clifford algebra with positive
signature, denoted by ${\cal C}(q+1,0)$,
\begin{equation}
\gamma _{\mu ik}\,\gamma _{\nu kj}+\gamma _{\nu ik}\,\gamma _{\mu
kj}=2\delta _{ij}\delta _{\mu \nu }\ .
\end{equation}
These Clifford algebras and some
of their properties are given in Table~\ref{Cliffp0} \cite{CliffordR}.
\begin{table}[htb]
\begin{center}
\begin{tabular}{||c|c|c|c|l||}\hline
$q$ &$q+1$ &${\cal C}(q+1,0)$& ${\cal D}_{q+1}$&${\cal S}_q(P,\dot P)$
\\ \hline &&&&\\[-3mm]
$-1$ & 0&$\Rbar$    &1         &$SO(P)$     \\
0    & 1&$\Rbar\oplus \Rbar $&1&$SO(P)\otimes SO(\dot P)$ \\
1    & 2&$\Rbar(2)$ &2         &$SO(P)$     \\
2    & 3&$\Cbar(2)$ &4         &$U(P)$     \\
3    & 4&$\Hbar(2)$ &8         &$U(P,\Hbar)\equiv USp(2P)$
\\ 4    & 5&$\Hbar (2)\oplus \Hbar (2)$&8&$USp(2P)\otimes USp(2\dot P)$\\
5    & 6&$\Hbar(4)$ &16&$U(P,\Hbar)\equiv USp(2P)$   \\
6    & 7&$\Cbar(8)$ &16&$U(P)$     \\
7    & 8&$\Rbar(16)$&16&$SO(P)$     \\
$n+7$ & $n+8$ & $\Rbar(16)\otimes{\cal C}(n,0)$&16 ${\cal D}_n$ &
as for $q+1=n$\\[1mm]
\hline
\end{tabular}
\end{center}
\caption{Real Clifford algebras ${\cal C}(q\!+\!1,0)$. Here
${\bf F}(n)$ stands for $n\times n$
matrices with entries over the field
$\bf F$, while ${\cal D}_{q+1}$ denotes the
real dimension of an irreducible representation of
the Clifford algebra. ${\cal S}_q(P,\dot P)$ is the metric
preserving group in the centralizer of the Clifford algebra in
the $(P+\dot P) {\cal D}_{q+1}$-dimensional representation.}
\label{Cliffp0}
\end{table}
The irreducible representations for a given $q$ are unique,
except when the Clifford module consists of a direct sum of two
factors.  As shown in Table~\ref{Cliffp0} this is the case for
$q=0$ mod 4, where there exist two inequivalent irreducible
representations.  This implies that, for $q\neq 0$
mod 4, the gamma matrices are unique once we specify the number of
irreducible representations. This number is denoted by $P$, so that
$L(q,P)$ determines the gamma matrices in this case, and thus
the $d_{ABC}$ coefficients.  For $q=0$ mod~4 the
representations $\gamma _\mu $ and $-\gamma _\mu $ are not
equivalent, and a reducible representation is characterized by
the multiplicity of each of these representations, $P$ and
$\dot P$. Of course, an overall sign change of all the gamma
matrices
can be absorbed into a redefinition of $h^\mu$ in \eqn{genC1}.
This is the reason why $L(4m,P,\dot P)=L(4m,\dot P,P)$.  If the
representation consists of copies of one version of the irreducible
representations, then we denote it by $L(4m,P)$.  The dimension of an
irreducible representation is denoted by ${\cal D}_{q+1}$ and given in
table~\ref{Cliffp0}.  We thus have $r=P\, {\cal D}_{q+1}$, or
$r=(P+\dot P)\, {\cal D}_{q+1}$, and thus\footnote{For  $L(q,P)$
one must take $\dot P=0$.}
\begin{equation}
n=3+q+(P+\dot P)\,{\cal D}_{q+1}\ .
\end{equation}

\subsection{Symmetries of the Clifford algebra.}

The isometry group of the various manifolds contains the linear
transformations of the
coordinates $h^A$ that leave the cubic polynomial \eqn{genC1}
invariant. A special subgroup of these invariance transformations
corresponds to the those that leave the Clifford
algebra invariant and preserve the metrics $\delta _{\mu \nu }$
and $\delta_{ij}$. In \cite{dWVP3} it was suggested that these
transformations are relevant for the homogeneous spaces that
occur in matter-coupled supergravity actions in 6 dimensions;
these couplings would then be determined by the Clifford algebra.

The symmetry transformations of the $\gamma$ matrices were
discussed in \cite{dWVP3}.  The symmetries of $\delta_{\mu\nu}$
form the rotation group $SO(q\!+\!1)$. They leave the $\gamma$ matrices
invariant when acting simultaneously on the spinor and vector
coordinates, labelled by $i$ and $\mu$, respectively.
On the spinor indices the rotations act according to the cover group.
Besides there can be additional invariances that act exclusively
in spinor space and commute with the gamma matrices and thus with
the corresponding representation of the Clifford algebra.  These
are the antisymmetric matrices $S_{ij}$ determined by
\begin{equation}
[\gamma _\mu ,S]=0
\ .\label{defSij}
\end{equation}
They are the metric-preserving elements of the centralizer of the
Clifford algebra representation, which are denoted
by\footnote{Both the group and the corresponding Lie
algebra will be denoted by ${\cal S}_q(P,\dot P)$.} ${\cal
S}_q(P,\dot P)$, and listed in Table~\ref{Cliffp0}.  The
invariance group of $\gamma_{\mu ij}$, $\delta_{ij}$ and
$\delta_{\mu\nu}$ is thus
\begin{equation}
SO(q+1)\otimes {\cal S}_q(P,\dot P)\ ,  \label{isod6}
\end{equation}
which acts according to
\begin{eqnarray}
\delta h^\mu &=&  A_{\mu\nu}\,h^\nu \ ,\nonumber\\
\delta h^i &=& \ft14A_{\mu \nu }\left( \gamma _\mu \gamma _\nu
\right)^i{}_j\,h^j+ S^i{}_j\,h^j\ , \label{symgamma}
\end{eqnarray}
where both $A_{\mu \nu }$ and $S_{ij}$ are anti-symmetric
matrices.
The dimension of the above invariance group equals
\begin{equation}
\half q(q+1) +\half e_q P(P+1) -P +\half e_q \dot P(\dot P+1)
-\dot P\ ,
\end{equation}
where $e_q$ equals 1, 2 or 4, depending on whether the Clifford
algebra is of real, complex or quaternionic type (see
Table~\ref{Cliffp0}). Hence $e_q=1$ for $q=0,1,7\bmod 8$, $e_q=2$
for $q=2,6\bmod 8$ and $e_q=4$ for $q=3,4,5\bmod 8$.

\subsection{Symmetries of the real space.}

The linear transformations of $h^A$ that leave (\ref{genC1})
invariant were already given in \cite{dWVP3}. They constitute the
isometries of the real special spaces and consist of
\eqn{symgamma} supplemented by (we rescaled some transformation
parameters for future convenience)
\begin{eqnarray}
\delta h^1 &=&- 2\lambda \,h^1 + 2\xi_i h^i   \ , \nonumber \\
\delta h^2 &=& \lambda \,h^2 -\zeta^i\, h^i+ \xi_\mu\,h^\mu \ ,
\nonumber \\
\delta h^\mu &=& \lambda \,h^\mu + \xi_\mu \,h^2 -
\zeta^j\,\gamma_{\mu ij}\,h^i \ ,\nonumber\\
\delta h^i &=&- {\textstyle{1\over 2}}\lambda \,h^i + \xi_i \,h^2
- \zeta^i\,h^1 + \xi_j \,\gamma_{\mu ij}\,h^\mu + \half\xi_\mu\,
\gamma_{\mu ij}\,h^j\ .
 \label{htrans}
\end{eqnarray}
The symmetries corresponding to the parameters $\zeta^i$ can only
exist when the tensor $\Gamma_{ijkl}=0$; this tensor is
defined by
\begin{equation}
\Gamma _{ijkl}\equiv\ft38\left[
\gamma _{\mu (ij}\,\gamma _{kl)\mu }-\delta _{(ij}\,
\delta_{kl)}\right] \ ,
\label{defGamma}
\end{equation}
and vanishes only when the corresponding \Ka\ and quaternionic spaces
are symmetric\footnote{ It does not vanish for $L(-1,P)$,  where only
the corresponding real space is symmetric, due to other
isometries than those discussed here \cite{sssl}. Henceforth we
only refer to  `symmetric spaces' when all three varieties of
the special spaces are symmetric.}. When this is not the case the
transformations \eqn{htrans} extend the number of symmetries  of
the previous subsection to
\begin{equation}
n-1 +\half q(q+1) +\half e_q \left(P(P+1) + \dot P(\dot
P+1)\right) - P -\dot P\ .
\end{equation}

\vspace{0.3cm}

We now clarify the algebra $\cal X$ corresponding to these
transformations.  First it is
easy to verify that generator $\underline{\lambda }$, associated
with the infinitesimal transformation proportional to $\lambda $,
extends the
Cartan subalgebra of \eqn{isod6} to the Cartan subalgebra of
$\cal X$. The rank of $\cal X$ is thus one unit higher than that
of \eqn{isod6}. Decomposing the
isometry algebra with respect to $\underline{\lambda }$ we find
\begin{equation}
{\cal X} = {\cal X}_{-3/2} + {\cal X}_0 + {\cal X}_{3/2} \ ,
\end{equation}
where ${\cal X}_{-3/2}$ contains the generators associated with
the parameters $\zeta^i$ (which is empty for the non-symmetric spaces),
${\cal X}_0$ consists of the generators associated with $\lambda
$, $\xi_\mu$,  $A_{\mu\nu}$ and $S_{ij}$, while ${\cal X}_{3/2}$
contains the generators corresponding to the parameters $\xi_i$.
We can simplify ${\cal X}_0$ to
\begin{equation}
{\cal X}_0 = so(1,1) \oplus so(q+1,1) \oplus {\cal S}_q(P,\dot P)\ ,
\end{equation}
where $so(1,1)$ corresponds to $\underline{\lambda }$ and the
$so(q+1,1)$ algebra consists of the $so(q+1)$ algebra discussed
in the previous subsection, combined with the generators
associated  with the infinitesimal transformations proportional
to $\xi_\mu$. It is convenient to decompose the fields $h^A$ into
$h^1$, $h^M$ and $h^i$, where $M= 2$ or $\mu$. The $so(q+1,1)$
infinitesimal transformations are then denoted by matrices
$A^M{}_N$ with $\xi _\mu = A^2{}_{\mu }=A^\mu{}_2$ and
$A^\mu{}_\nu=A_{\mu\nu}$, with $A_{MN}= \eta_{MP}\,A^P{}_N=
-A_{NM}$, where the metric $\eta= \hbox{diag}\,(-1,1,\ldots,1)$.

On the spinor fields $h^i$ the group $SO(q+1,1)$ act as indicated
in \eqn{htrans} (i.e. it transforms as a
spinor in de Sitter space). From a more formal point of view,
this can be understood as follows. The spinor fields do not allow
a realization of ${\cal C}(q+1,1)$, but its subset consisting of
the even elements, ${\cal C}^+(q+1,1)$, is always isomorphic with
${\cal C}(q+1,0)$, such that $SO(q+1,1)$ can act on $h^i$. To see
this in more detail, let us first double
the representation space and realize a (not necessarily
irreducible) representation of the  Clifford algebra ${\cal
C}(q+1, 1)$. This representation thus acts on $2{\cal D}_{q+1}$
coordinates, which we decompose into an equal number of
components with upper and with lower indices, $(\psi^i, \psi_i)$.
As $\psi^i$ and $\psi_i$ transform in general according to
inequivalent spinor representations of $SO(q+1,1)$, there is no
invariant metric in spinor space that allows one to raise and
lower spinorial indices, so that $\psi^i$ and $\psi_i$ remain
independent. On this basis ${\cal C}(q+1, 1)$ is realized by the
gamma matrices
\begin{equation}
\gamma_M= \pmatrix{0 & \gamma_{M\,ik}\cr
       \noalign{\vskip 1mm}
       \gamma_M{}^{jl} & 0 \cr }\ ,
\end{equation}
with $\gamma_{\mu\,ij}=\gamma_\mu{}^{ij}$ equal to the gamma
matrices introduced above, and $ \gamma_2{}^{ij}= -\gamma_{2\,
ij}= \delta_{ij}$. The generators of $SO(q+1,1)$ in the spinor
representation are
\begin{equation}
 \gamma_{MN}{}^i{}_j= \gamma _{[M}{}^{ik}\,\gamma _{N]\,kj}\ ,
\end{equation}
so that
\begin{equation}
\gamma _{\mu \nu }{}^i{}_j=\gamma _{[\mu }{}^{ik}\,\gamma_{\nu
]kj}\ ;\qquad
\gamma _{2\mu }{}^i{}_{j}=-\gamma _{\mu 2}{}^i{}_{j}=\gamma _\mu{}^{ij} \ .
\end{equation}

With this notation the only non-zero components of the tensor
$d_{ABC}$ are
\begin{equation}
d_{1MN}= -\eta _{MN}\ ;\qquad d_{Mij}=\gamma _{Mij}\ ,
\label{dSOq+11}
\end{equation}
where the coordinates $h^i$ are assigned upper indices. Assigning
upper indices to the generators $\underline{\xi}{}^i$ and lower
indices to $\underline{\zeta}{}_i$ (so that the corresponding
parameters are written as $\xi_i$ and $\zeta^i$, respectively),
the commutators with the $so(q+1,1)$ generators are
\begin{equation}
[\underline{A}_{MN},\underline{\xi }^i]=-\half \gamma
_{MN}{}^i{}_{j}\,\underline{\xi }^j \ ; \qquad
[\underline{A}_{MN}, \underline{\zeta}{}_i]=
\half\underline{\zeta}{}_j \,\gamma_{MN}{}^j{}_{i} \ ,
\end{equation}
and the combined transformations of \eqn{symgamma} and \eqn{htrans}
are incorporated in the matrix ($A=(1,M,i)$ and $B=(1,N,j)$)
\begin{equation}
\tilde B^A{}_{\!B}=\pmatrix{-2\lambda & 0& 2\xi _j \cr
    \noalign{\vskip 2mm}
         0&\lambda\,\delta ^M_N+A^M{}_{N} & -\zeta^k\,\gamma^M{}_{kj} \cr
     \noalign{\vskip 2mm}
         -\zeta^i & \gamma _N{}^{ik}\xi_k & -\half \lambda
    \delta^i_j +\ft14 A^{PQ}\left( \gamma _{PQ}\right)^i{}_{j}
    +S^i{}_{j}\cr}\ .
\label{tilBhom}
\end{equation}

To find the isotropy group, consider points with $h^\mu =h^i=0$. Then
$h^1$ depends on $h^2$ because of \eqn{dhhh1}, so that one is
left with a one-dimensional subspace. For these points it is easy
to show that the metric (defined in \eqn{5dLagr}) is negative definite as
required. None of these points is left invariant by the
symmetries \eqn{htrans} (this is not so for the symmetric spaces
where the isotropy group contains transformations associated with a
linear combination of $\xi_i$ and $\zeta^i$), so that the
(compact) isotropy group equals
\begin{equation}
H= SO(q+1) \otimes {\cal S}_q(P,\dot P)\ .
\end{equation}

The $(n-1)$-dimensional solvable subalgebra of the non-compact
isometry group $G$ consists of two parts, namely the $r$
generators contained in  ${\cal X}_{3/2}$, and the $q+2$
generators belonging to the solvable subalgebra of ${\cal X}_0$;
those are the generator $\underline{\lambda }$ and the
$q+1$ generators belonging to the solvable subalgebra of $so(q+1,
1)$ (see appendix~\ref{app:iwasawa}). The rank of the $so(q+1,1)$
equals  1 (for $q\geq 0$; for $q=-1$ the algebra is empty, so
that the rank is 0), so
that the rank of the full solvable algebra equals 2 (or 1 for
$q=-1$). Obviously the real homogeneous spaces have the form of
${SO(q+1,1)\over SO(q+1)}$ supplemented with $r+1$ coordinates
associated with the rigid motions parametrized by $\lambda $
and $\xi _i$, which act as translations and transform
linearly under $SO(q+1,1)\otimes {\cal S}_q(P,\dot P)$. For
symmetric spaces the situation is different as one has to take
into account the extra isometries contained in ${\cal X}_{-3/2}$.

Note that for $q=-1$ we have ${\cal X}_0= so(1,1)\oplus so(P)$.
Then the $r=P$ generators of ${\cal X}_{3/2}$ and the additional
isometries found in \cite{sssl} extend this algebra to $so(P+1,
1)$, and one finds a symmetric space as indicated in
Table~\ref{homsp}.

{}For $q=0$ we have ${\cal X}_0= so(1,1)\oplus so(1,1)\oplus
so(P)\oplus so(\dot P)$. The generators $\underline{\lambda }$ and
$\underline{\alpha }$ (the one corresponding to $\xi_\mu $), can
be recombined such that one linear combination commutes with the
$P$ generators in ${\cal X}_{3/2}$ that transform under $SO(P)$,
and the other linear combination commutes with the remaining
$\dot P$ ones transforming under $SO(\dot P)$. The space then
becomes a local product of a
$(P+1)$- and a $(\dot P+1)$-dimensional space (each of rank 1).
The isometry algebra is a contraction of $so(P+1,1)\oplus so(\dot
P+1,1)$. These spaces were denoted by $Y(P,\dot P)$ in Table~\ref{homsp}.
If $q=\dot P=0$ the space is symmetric and one must include the
$P$ generators contained in ${\cal X}_{-3/2}$. This then leads to
the isometry algebra $so(P+1,1)\oplus so(1,1)$, as exhibited in
Table~\ref{homsp}.

\subsection{Symmetries of the \Ka\ space.}

The special \Ka\ spaces are based on the function $F(X)$, which
for the `very special' spaces takes the form
\begin{equation}
F(X)= {3i\over X^0} \left\{-\eta_{MN}\,X^MX^NX^1 + \gamma_{M\,ij}
\, X^MX^iX^j\right\} \ . \label{Fdhom}
\end{equation}
The previous isometry algebra is now extended with the
transformations corresponding to $\beta $, $b^A$ and the solutions of
\eqn{4conda}. So we first have to solve the latter. As the spaces are
homogeneous, we may do so at any point in the domain. It is
convenient to consider points in the subspace $x^\mu =x^i=0$, so
that only $x^1$ and $x^2$ are
non-vanishing (and possibly the real parts of $z^A$, but they never
appear). The metric is diagonal in this parametrization and is given
by
\begin{equation}
g_{11}=-\left( x^1\right) ^{-2}\ ;\quad
g_{22}=-2\left( x^2\right) ^{-2}\ ;\quad
g_{\mu \nu }=-2\delta_{\mu\nu }\left( x^2\right) ^{-2}\ ;\quad
g_{ij}=-2\delta_{ij}\left(x^1x^2\right) ^{-1}\ .
\end{equation}
As $dxxx=3x^1\left( x^2\right) ^2$, the domain of positivity is
$x^1,\,x^2>0$.
The non-zero components of $C^{ABC}$ are (cf. \eqn{defCABC})
\begin{equation}
C^{122}=\ft34\ ;\qquad C^{1\mu \nu }=-\ft34\delta^{\mu \nu }\ ;\qquad
C^{2ij}=-\ft38\delta^{ij}\ ;\qquad C^{\mu ij}=\ft38\gamma _{\mu ij}
\label{Chom}\end{equation}
or, in $so(q+1,1)$-covariant notation,
\begin{equation}
C^{1MN}= -\ft34\eta ^{MN}\ ;\qquad C^{Mij}=\ft38\gamma^{Mij}\ .
\end{equation}
The curvature tensor, which appears in the commutation rules of
the isometries, follows directly from the above result. Its non-zero
components in the points $x^\mu =x^i=0$ are
\begin{eqnarray}                &&
R^1{}_{11}{}^1=-2\ ;\qquad R^1{}_{1j}{}^i=-\delta ^i_j\ ,
\nonumber\\
&&R^1{}_{ij}{}^M=-\gamma ^M{}_{ij}\ ;\qquad R^i{}_{1M}{}^j=-\ft12
\gamma_M{}^{ij}\ ,\nonumber\\
&& R^P{}_{\!MN}{}^Q=-2\delta _M^{(P}\,\delta _N^{Q)}+\eta _{MN}\,
\eta^{PQ}\ ,\nonumber\\
&& R^N{}_{\!Mi}{}^j=-\half \gamma _M{}^{jk}\,\gamma^N{}_{ki}\ ,\nonumber\\
&& R^k{}_{ij}{}^l=-2\delta _i^{(k}\,\delta _j^{l)}+\half
\gamma_{Mij} \,\gamma ^{Mkl} \ .
\end{eqnarray}
Furthermore, a straightforward calculation gives for the non-zero
components of $E^E_{ABCD}$,
\begin{equation}
E^1_{ijkl}=2\Gamma _{ijkl}\ ;\qquad
E^i_{M jkl}=\Gamma _{jklm}\gamma_M{}^{ mi}\ , \label{Ehom}
\end{equation}
where $\Gamma_{ijkl}$ was defined in \eqn{defGamma}. We consider
the non-symmetric spaces where this tensor is non-zero. Already
in \cite{dWVP3} we concluded that  there are no non-trivial
solutions of $\Gamma _{ijkl}\,a_l=0$ in that case. Therefore there
are only hidden \Ka\ symmetries associated with the parameters
$a_\mu $ and $a_2$.

According to \eqn{calW} the isometry algebra ${\cal W}$ can be
decomposed into three eigenspaces of $\underline{\beta }$; ${\cal
W}_{-2/3}$ contains the transformations corresponding to $a_M$,
${\cal W}_0={\cal X}\oplus \underline{\beta}$, and ${\cal
W}_{2/3}$ contains the generators associated with the parameters
$b^A$. The total number of isometries is now
\begin{equation}
2(n+1)+\half q(q+3)+\half e_q\left( P(P+1)+\dot P(\dot
P+1)\right) -P-\dot P\ .
\end{equation}
The parameters $b^A$ decompose with respect to $SO(q+1,1)$ into
$b^1$, $b^M $ and $b^i$. Most of the commutation rules are
indicated in
Table~\ref{rootsdKa}.
\begin{table}[htf]
\caption{Roots of the isometries of the non-symmetric
homogeneous very special \Ka\ spaces.}
\label{rootsdKa}\begin{center}\begin{tabular}{||l|ccrr|rr||}\hline
generator &${\cal S}_q$& $so(q+1,1)$ &\underline{$\lambda $}  &
$\underline{\beta }$ &$\underline{\beta }-\ft13 \underline{\lambda }$&
$\ft23\underline{\lambda }+\underline{\beta }$
\\[2mm] \hline
$\underline{b}_1$&0& 0   & 2   & $\ft23$&0&2\\[2mm]
$\underline{\xi}^i$& $v$ & $s$ & $\ft32$ & 0 &$-\ft12$ & 1\\[2mm]
$\underline{b}_i$  &$v$& $\bar s$ & $\ft12$ &$\ft23$&$\ft12$ & 1\\[2mm]
$\underline{a}^M$&0& $v$ & 1 &$-\ft23$&$-1$&0\\[2mm]
$\underline{b}_M$&0& $v$ & $-1$ & $\ft23$&1&0\\[2mm]
 \hline
\end{tabular}\end{center}\end{table}
The ${\cal S}_q$ representation denoted by $v$ denotes the
vector (or defining) representation. The $so(q+1,1)$
representation denoted by $v$, $s$ and $\bar s$
denotes the vector, spinor and conjugate spinor representation,
respectively. Hence we have
\begin{eqnarray}
[\underline{A}_{MN},\,\underline{a}^P]&=&
2\underline{a}_{[M}\,\delta ^P_{N]} \ ,\nonumber\\{}
[\underline{A}_{MN},\,\underline{\xi }^i]&=&-\half \gamma
_{MN}{}^i{}_{j}\,\underline{\xi }^j \ , \nonumber\\{}
[\underline{A}_{MN},\,\underline{b}_i]&=&\half\underline{b}{}_j \,\gamma
_{MN}{}^j{}_{i} \ .
\end{eqnarray}
To calculate the last commutator of \eqn{comKad} we can use the
curvature tensor as determined above for points in the
two-dimensional complex subspace,
because we know that the components $R^M{}_{\!\!AB}{}^{\!C}$ are
related to the hidden symmetries parametrized by $a_M$, and
therefore constant. In this way we obtain
\begin{eqnarray}
[\underline{\xi }^j,\,\underline{b}_M]&=&-\underline{b}_i\,\gamma
_M{}^{ij}\ ,\nonumber\\{}
[\underline{\xi }^j,\,\underline{b}{}_i]&=&-2\delta ^j_i \,
\underline{b}{}_1 \ ,\nonumber\\{}
[\underline{b}_M,\,\underline{a}^N]&=&\delta ^N_M
(\underline{\beta }-\ft13 \underline{\lambda
})-\underline{A}^N{}_{M}\ , \nonumber\\{}
[\underline{b}{}_i,\,\underline{a}^N]&=& -\half\gamma ^N{}_{ij} \,
\underline{\xi }^j \ .
\end{eqnarray}
Note that $\underline{b}{}_1$ commutes with the generators
discussed above, as well as with the $so(q+1,1)$ generators.
Furthermore, it is easy to verify that the generators
$\underline{A}{}^M{}_{\!N}$, $\underline{a}^M$,
$\underline{b}{}_M$ and $\underline{\beta}
-\ft13\underline{\lambda }$ define the algebra $so(q+2,2)$.
Extending the indices $M$ with two more values, which we indicate
by $a$ and
$b$, we may define $\underline{A}_{aM}\equiv \underline{a}_M$,
$\underline{A}_{bM}\equiv \underline{b}_M$, and
$\underline{A}_{ab}=-\underline{A}_{ba}=\underline{\beta
}-\ft13\underline{\lambda }$. The invariant metric $\eta_{MN}$ is
then extended to an $SO(q+2,2)$ invariant metric by including
$\eta_{ab}=\eta _{ba}=-1$.

According to the above commutation relations, the spinor
representation of $SO(q+2,2)$ acts on the generators
$(\underline{\xi }^i,\underline{b}{}_i)$. This can be understood
from the equivalence relations ${\cal C}^+(q+2,2)\simeq{\cal
C}(q+2,1)\simeq {\cal C}(q+1,0)\otimes \Rbar (2)$.

The linear
combination of $\underline{\lambda}$ and $\underline{\beta }$
that commutes with $so(q+2,2)$ is $\underline{\lambda
}'=\ft23\underline{\lambda} +\underline{\beta }$. The grading of the
full isometry algebra with respect to $\underline{\lambda}'$ is
as follows,
\begin{eqnarray}
&&{\cal W}={\cal W}'_0 \oplus  {\cal W}'_1 \oplus  {\cal W}'_2\ ,
\nonumber\\
&&{\cal W}'_0= \underline{\lambda }'\oplus so(q+2,2)\oplus {\cal
S}_q(P,\dot P) \ ,\nonumber\\
&&{\cal W}'_1= \underline{\xi }^i \oplus \underline{b}_i =(1, s, v)\
,\nonumber\\
&&{\cal W}'_2= \underline{b}_1=(2,0,0)\ , \label{homWroots}
\end{eqnarray}
where, for ${\cal W}'_1$ and ${\cal W}'_2$, we indicated the
representations with respect to the three subalgebras of ${\cal W}'_0$.
Note that because of the equivalence of the Clifford algebras
indicated above, ${\cal S}_q(P,\dot P)$ is also the centralizer
of the ${\cal C}^+(q+2,2)$ in the
relevant representation. In this picture, the generators of grading 0 form
again a reductive algebra, while the other generators have positive
grading~: at grading 1 the generators constitute a direct product
of a spinor representation of $so(q+2,2)$ with a vector
representation of ${\cal S}_q$; at grading 2 there is just a
singlet generator. No generators with negative grading occur.
This is different for the symmetric spaces, where we have
additional isometries associated with the parameters $\zeta^i$,
$a_1$ and $a_i$. They decompose into generators
$(\underline{\zeta}{}_i,\underline{a}^i)$ at grading $-1$ and
$\underline{a}^1$ at grading $-2$, so that the isometry algebra
is semisimple.

Let us briefly consider the simplest cases. For the homogeneous spaces
related to $L(-1,P)$, the reductive algebra ${\cal W}'_0$ is
$\underline{\lambda }'\oplus so(2,1)\oplus so(P)$. The generators in
${\cal W}'_1$ form a $(1,2,P)$ representation of this algebra.
For $q=0$, we have the so-called $K(P,\dot P)$ spaces and
\begin{eqnarray}
{\cal W}'_0 &=& \underline{\lambda }'\oplus so(2,2)\oplus so(P) \oplus
so(\dot P) \nonumber\\
&=& \underline{\lambda }'\oplus \big( so(2,1)\oplus so(P)\big) \oplus
\big(so(2,1)\oplus so(\dot P)\big)\ .
\end{eqnarray}
The generators in ${\cal W}'_1$ decompose into two separate
representations, $(1,2,P,0,0)$ and $(1,0,0,2,\dot P)$.
Because of the presence of $\underline{\lambda }'$ and
$\underline{b}_1$ in the isometry algebra, this does not lead to
a factorization of the corresponding \Ka\ space. The latter two
generators form a so-called canonical subalgebra,
characteristic for the K\"ahlerian algebras \cite{Aleks,Cecotti}.

{}From the transformation rules
\eqn{ztrans}, it is obvious that the coordinates $z^A$ do not transform
linearly under the isometry group. For the real spaces, discussed
in the previous subsection, the transformations were realized
linearly on the fields $h^A$, which are, however, subject to the
constraint \eqn{dhhh1}, and the cubic polynomial \eqn{genC1} is
invariant under the isometry transformations. For the \Ka\
spaces there are the corresponding fields $X^I$, which do not
transform linearly either as is
shown in \eqn{scaltra}. Furthermore the homogeneous function
$F(X)$ given in \eqn{Fdhom} is not invariant under the full
isometry group, and neither is it invariant under the isotropy group.
The latter group coincides with the compact subgroup associated with
${\cal W}'_0$ and equals
\begin{equation}
H= SO(q+2)\otimes SO(2) \otimes {\cal S}_q(P,\dot P)\ .
\end{equation}
The $SO(q+2)$ group is associated with the compact subgroup of
$SO(q+1,1)$ and linear combinations of the generators
$\underline{a}^\mu$ and $\underline{b}{}_\mu$; the $SO(2)$ group
is generated by a linear combination of $\underline{a}^2$ and
$\underline{b}{}_2$ (and not by linear combinations of
$\underline{\lambda}$ and $\underline{\beta}$). In general it is
not possible to use a symplectic reparametrization (cf.
appendix~\ref{app:symrepar}) to bring
$F(X)$ into a different form that is manifestly invariant under
the isotropy group $H$.

\subsection{Symmetries of the quaternionic space.}
The symmetries of the generic quaternionic very special spaces are
summarized by table~\ref{symd}. In the previous subsection, we
have already identified the symmetries corresponding to $\tilde
B^A{}_{B}$ and $a_A$. As we know that the condition for the
existence of the $a_A$ symmetries coincides with the condition
for the existence of the $\hat \beta _A$ symmetries, there are
thus $q+2$ solutions associated with the parameters $\hat \beta
_M$. We are left with the $\hat\alpha{}^A$ symmetries, whose
existence is governed by condition \eqn{condhatal}.
{}From \eqn{alphabebe} we derive at once, using  \eqn{Chom}, that
there is a symmetry associated with $\hat \alpha^1$. Using the
$E$ coefficients in \eqn{Ehom} establishes that the
non-symmetric spaces have no other symmetries. Another
approach is to consider \eqn{dSOq+11} and to note that
additional symmetries associated with $\hat \alpha{}^A$ with
$A\not= 1$, would lead to $\hat \beta_A$ symmetries of the form
$\hat\beta_A=d_{ABC}\,\hat\alpha{}^B$ that are not realized.

Hence the isometry algebra of the homogeneous quaternionic spaces
consist of the algebra \eqn{homWroots} of the corresponding \Ka\
spaces extended with $\hat\beta{}^M$ and $\hat\alpha{}_1$. Using
\eqn{comduala}, \eqn{bigmatd} and \eqn{tilBhom}, we can group the
generators into representations of  $\underline{\epsilon}{}_0
\oplus \underline{\lambda}'\oplus so(q+2,2)\oplus {\cal S}_q(P,
\dot P)$. The results are listed in Table~\ref{rootsqhom},
\begin{table}[htf]
\caption{Roots of the isometries of the non-symmetric
homogeneous very special quaternionic spaces}
\label{rootsqhom}\begin{center}\begin{tabular}{||l|ccrr|rr||}\hline
generator &${\cal S}_q$& $so(q+2,2)$ & $\underline{\lambda}'$  &
$\underline{\epsilon}{}_0$ &$\underline{\lambda}'-2
\underline{\epsilon}{}_0$&
$\underline{\lambda}'+2\underline{\epsilon}{}_0$
\\[2mm] \hline
$\underline{\epsilon}{}_+$&0&0&0&1&$-2$&2\\[2mm]
$(\underline{\alpha}{}_1,\underline{\beta }^M,\underline{\beta }^0)$
&0& $v$&1&$\ft12 $&0&2 \\[2mm]
$\underline{b}_1$&0& 0 & 2& 0&2&2\\[2mm]
$(\underline{\alpha}{}_i,\underline{\beta}{}^i)$ & $v$&$s$ &0&$\ft12$&$-1$&1
\\[2mm]
$(\underline{\xi}^i,\underline{b}{}_i)$& $v$&$\bar s$ &1&0&1&1\\[2mm]
$(\underline{\hat \alpha}{}_1,\underline{\hat \beta}{}^M,\underline{\hat
\beta}{}^0)$&0& $v$&1&$-\ft12$& 2&0 \\[2mm]
$(\underline{\alpha }_0,\underline{\alpha }_M,\underline{\beta }^1)$
&0& $v$&$-1$&$\ft12$&$- 2$&0 \\[2mm]
 \hline
\end{tabular}\end{center}
\end{table}
where, in the first column, the generators that constitute an
$so(q+2,2)$ representation are ordered according to their weight
under the $so(q+2,2)$ generator
$\underline{\beta}-\ft13\underline{\lambda }$. For the vector
representation these weights are $(-1,0,1)$, and for the spinor
representation $(-\ft12,\ft12)$.

{}From the table we see immediately that  $so(q+2,2)$ can be
extended with $(\underline{\hat \alpha}{}_1,
\underline{\hat\beta}{}^M, \underline{\hat\beta}{}^0)$,
$(\underline{\alpha}{}_0,\underline{\alpha}{}_M,
\underline{\beta}{}^1)$  and
$\underline{\lambda}'-2\underline{\epsilon }_0$ to $so(q+3,3)$.
This can be checked by combining \eqn{commalhal} and
\eqn{pardualcom}. The remaining generators can be grouped into
$so(q+3,3)$ representations. Denoting the linear combination of
$\underline{\epsilon}_0$ and $\underline{\lambda}'$ that commutes
with $so(q+3,3)$, by $\underline{\epsilon }'\equiv
2\underline{\epsilon}_0+\underline{\lambda}'$, we find a
decomposition of the isometry algebra as in the previous
subsection~:
\begin{eqnarray}
{\cal V}&=& {\cal V}'_0 +{\cal V}'_{1} + {\cal V}'_2 \ ,\nonumber\\
{\cal V}'_0 &=&\underline{\epsilon }'\oplus so(q+3,3)\oplus {\cal
S}_q(P,\dot P)\ , \nonumber\\
{\cal V}'_1 &=& (\underline{\xi}^i,\underline{b}_i)
\oplus(\underline{\alpha}{}_i,\underline{\beta}{}^i) =  (1,s,v) \ ,
\nonumber\\
{\cal V}'_2 &=& \underline{\epsilon}{}_+\oplus
(\underline{\alpha}{}_1,\underline{\beta}{}^M,
\underline{\beta}{}^0) \oplus \underline{b}{}_1 = (2,v,0) \ ,
\end{eqnarray}
where we indicated the representation of ${\cal V}'_1$ and ${\cal
V}'_2$ according to the three subalgebras of ${\cal V}'_0$. The
above result fully determines the generic structure of the isometry
algebra for the (non-symmetric) homogeneous
quaternionic spaces. The isotropy group is the maximal compact
subgroup of the isometry group, which equals
\begin{equation}
H= SO(q+3)\otimes SO(3)\otimes {\cal S}_q(P,\dot P)\ .
\end{equation}
For the symmetric spaces the isometry algebra is extended with
additional generators, with negative eigenvalues of
$\underline{\epsilon}'$,
\begin{eqnarray}
{\cal V}'_{-1} &=& (\underline{a}^i,\underline{\zeta}_i)
\oplus(\underline{\hat\alpha}{}_i,\underline{\hat\beta}{}^i) =
(-1,s,v) \ ,\nonumber\\
{\cal V}'_2 &=& \underline{\epsilon}{}_-\oplus
(\underline{\hat\alpha}{}_0,\underline{\hat \alpha}{}_M,
\underline{\hat\beta}{}^1) \oplus \underline{a}{}^1 = (-2,v,0)\ .
\end{eqnarray}

\vspace{ 1cm}

\noindent{ \bf Acknowledgements}\vspace{0.3cm}

We thank S.~Schrans, who contributed to this work at an initial
stage, and S.~Ferrara, J.~Louis, V.I.~Ogievetsky and W.~Troost
for valuable
discussions.  This work was carried out in the framework of the
project "Gauge theories, applied supersymmetry and quantum
gravity", contract SC1-CT92-0789 of the European Economic
Community.

\appendix
\section{Normal spaces}
\label{app:iwasawa}
\setcounter{equation}{0}

For homogeneous spaces the isometries act transitively on
the manifold so that every two points are related by an element
of the isometry group. The orbit swept out by the action of the isometry
group $G$ from any given point is (locally) isomorphic to the
coset space $G/K$, where $K$ is the isotropy group of that point.
For non-compact homogeneous spaces where $K$ is the maximal compact
subgroup of $G$, there exists a solvable subgroup that acts
transitively and whose dimension is equal to the dimension of the
space.
Such spaces are called {\em normal}.
This solvable algebra $s$ determines the coset completely, i.e.
$\frac{G}{K}=e^s$. We will show how to obtain $s$ for any coset.

The construction of the solvable algebra
follows from the Iwasawa decomposition of the group,
\begin{equation}
G= K\,A\,N \ ,
\end{equation}
where $K$ is the maximal compact subgroup; the remaining factor
is then decomposed into its maximal Abelian subgroup $A$ and a
nilpotent group $N$. The solvable subgroup is $F=A\,N$.

For normal spaces the compact subgroup can be divided out
according to the above decomposition so that  one
is left with the group space associated with $F$. To see how this
works, we first consider the example of $SO(n,1)/SO(n)$, and show how this
coset is indeed the exponential of a solvable algebra. Then we
explain the Iwasawa decomposition for the algebra. Finally we
illustrate how this decomposition works for the example of $so(n,
1)$.

One decomposes the generators of $SO(n,1)$ using an off-diagonal
metric decomposed according to $(n+1) = 1 \oplus (n-1)\oplus 1$,
\begin{equation}
\eta = \pmatrix{0 &  0   & 1 \cr
                0 & \unity  & 0 \cr
                1 &  0   & 0 \cr}   \,.      \label{etapm}
\end{equation}
The generators of $SO(n,1)$ are subject to the condition
$t^{\rm T} =- \eta\, t \,\eta$ and can be
decomposed as follows,
\begin{equation}
t\propto \pmatrix{\alpha  &  -\xi^{\rm T}    & 0       \cr
                   \noalign{\vskip 1mm}
                   \zeta  & so(n-1)          &  \xi    \cr
                   \noalign{\vskip 1mm}
                   0      &  -\zeta^{\rm T}  & -\alpha \cr}   \,.
\label{tdecomp}
\end{equation}
Here $\xi$ and $\zeta$ are $(n-1)$-dimensional real vectors. The
$so(n-1)$ generators together with the remaining $n-1$ compact
generators characterized by $\xi=\zeta$ define the maximal compact
subalgebra $so(n)$. Observe that the vector left invariant under
$SO(n)$ is equal to $(1,0,\ldots 0, -1)$. Acting on this vector
with $SO(n,1)$ generates an orbit isomorphic to $SO(n,1)/SO(n)$
as well as to $F =\exp s$, where $s$ is the solvable algebra
associated with the generators parametrized by $\alpha$ and
$\xi$. It is not difficult to give this group explicitly in terms
of these parameters,
\begin{equation}
{\rm e}^{s(\alpha,\xi)} = \pmatrix{\displaystyle{\rm e}^\alpha &
\displaystyle {1-{\rm e}^\alpha\over \alpha}\,\xi^{\rm T}
&\displaystyle {1-\cosh \alpha \over \alpha^2}\, \xi^2 \cr
\noalign{\vskip3mm}
0 & 0 & \displaystyle {1-{\rm e}^{-\alpha}\over \alpha}\,\xi \cr
\noalign{\vskip5mm}
0 & 0 & {\rm e}^{-\alpha}\cr}
\end{equation}
The rank of the coset is defined as the rank of the solvable
algebra, which in this case is equal to 1. The Cartan subalgebra
consists of the generator associated with the parameter $\alpha$.
\vspace{0.3cm}

We now present the general procedure (see e.g. \cite{Helg}) for the
Iwasawa decomposition of a non-compact algebra
\begin{equation}
g=k\oplus p \ ,
\end{equation}
such that \begin{equation}
u=k\oplus ip
\end{equation}
is compact. We will thus explain how to find the solvable algebra $s$.
First, choose a maximal number of commuting elements from
$p$, and denote them by $h_p$. This set can be extended to a
Cartan subalgebra (CSA) $h$ by including suitable elements from
$k$. We order the elements in the CSA by putting the elements of
$h_p$ in front, and then define the set of positive roots $\Delta ^+$.
Now $n$ is the subset of $\Delta ^+$ defined by those elements which
are not completely in $k$. (More mathematically~: define an
automorphism $\theta $ such that $\theta (k)=k$ and $\theta (p)=-p$,
then $n$ is the subset of $\Delta ^+$ consisting of elements $x$ such
that $\theta (x)\neq x$). Equivalently, $n$ consists of
the roots which are strictly positive with respect to $h_p$ only.
Then
\begin{equation}
s=h_p\oplus n
\end{equation}
is a solvable algebra, which has the same dimension as $p$ (see
\cite{Helg}) and $e^s$ is isomorphic with the coset $g/k$.
The dimension of the Cartan subalgebra of $s$ equals the {\it rank} of the
homogeneous space.

Consider now the example of $so(n,1)$. Let us define the algebra by using
the vector representation. Indices $M,N,\ldots,$ run over the values
$0,1,\ldots,n$. Indices $\mu ,\nu$ run from 1 to $n$. Furthermore
we will also use indices $m,n,\ldots$ running from 2 to $n$. We
define transformations
\begin{equation}
\delta z^M = A^M{}_Nz^N\ ,
\end{equation}
and raise or lower indices with the metric $\eta ^{MN}=
\hbox{diag}\,(-1,1,\ldots,1)$. Then $A_{MN}$ is
antisymmetric. The generators $\underline{A}$ are defined by $\delta =\half
A^M{}_{N}\,\underline{A}^N{}_M$. The compact ones ($\underline{A}^\mu {}_{
\nu }$) are antisymmetric operators, while the non-compact ones
($\underline{A}^0{}_{\mu }$) are symmetric. Their action is
\begin{equation}
\underline{A}_{MN}z^N=\delta ^P_N z_M -\delta ^P_M z_N\ .
\end{equation}
The commutation relations
\begin{equation}
[\underline{A}_{MN},\,\underline{A}_{PQ}]=\underline{A}_{MQ}\,\eta _{NP}
-\underline{A}_{NQ}\,\eta _{MP}-\underline{A}_{MP}\,\eta _{NQ}+
\underline{A}_{NP}\,\eta _{MQ}
\end{equation}
imply that there are no mutually commuting
non-compact generators. The space is therefore of rank 1,
and we can choose $\underline{\alpha }\equiv \underline{A}^0{}_1$ as the
generator in $h_p$ (the full $h$ includes also the Cartan subalgebra
of $so(n-1)$). Now we have to diagonalize the commutator of $h_p$
with the other generators. This requires the linear combinations
\begin{equation}
\underline{A}^\pm {}_m=\frac{1}{\sqrt{2}}\left(
\underline{A}^0{}_m\pm \underline{A}^1{}_m\right)\ .
\end{equation}
On this basis the metric becomes equal to the metric \eqn{etapm}.
The generators $\underline{A}^+{}_m$ have a positive
root with respect to
$\underline{\alpha} $, and correspond to the parameters $\xi_m$ in
\eqn{tdecomp}. Thus we find  that the $n-1$ generators
$\underline{\xi}{}_m$ and $\underline{\alpha }$ generate the
$n$-dimensional coset $so(n,1)/so(n)$; the only non-zero commutators
of the solvable algebra are
\begin{equation}
[\underline{\alpha },\,\underline{\xi }{}_m]=\underline{\xi }{}_m\ .
\end{equation}

\setcounter{equation}{0}
\section{Useful formulae}   \label{app:formN2}
We collect here several formulae which are useful in calculations.
The matrices in \eqn{defmat} have simple contractions with
the vector~$X$
\begin{eqnarray}
X^I\cM_{I\bar J}&=&0\nonumber\\
X^I\cN_{IJ}&=&-\frac{1}{4}X^IF_{IJ}=-\frac{1}{4}F_J\nonumber\\
(XN\bar X)(XN)_I&=&-(XNX)\bar X^J(\cN +\bar \cN)_{JI}.
\end{eqnarray}
For the inverse matrices there are the following relations (the
matrix ${\cal M}^{-1\, \bar AB}$ is the inverse of ${\cal M}_{A\bar B}$ as
$n\times n$ matrix)
\begin{eqnarray} &&{\cal M}^{-1\,
\bar AB}=(zN\bar z)^{-1} g^{\bar A B}\nonumber\\ && \hspace{1.5cm} =
\big(N^{-1}\big){}^{\bar
A B} -\big(N^{-1}\big){}^{\bar A0}\, z^B -\bar z{}^{\bar A}\,
\big(N^{-1}\big){}^{0B} + \big(N^{-1}\big){}^{00} \, \bar
z{}^{\bar A}\, z{}^{B}\nonumber\\
&&(N^{-1})^{IJ}=(\cN +\bar \cN)^{-1\ IJ}+\frac{X^I\bar X^J+\bar
X^IX^J}{XN\bar X}\nonumber\\
&&(\cN +\bar \cN)^{-1\ IK}N_{KJ}=(N^{-1})^{IK}\cM_{K\bar
J}-\frac{X^I(N\bar X)_J}{XN\bar X}\nonumber\\
&&\cN\frac{1}{\cN+\bar \cN}\bar
\cN=\bar \cN\frac{1}{\cN+\bar \cN}\cN= \cN-\cN\frac{1}{\cN+\bar \cN} \cN
\end{eqnarray}
We have defined in \cite{BEC} also $\left( N^{-1}\right){}^{00}=\Delta
^{-1}$ and the matrix $\left( n^{-1}\right){}^{AB}$ as the inverse of
$N_{AB}$. We then have
\begin{equation}
\Delta =zN\bar z - zN\left( n^{-1}\right) N\bar z=N_{00}-N_{0A}\left(
n^{-1}\right){}^{AB}N_{B0}\ .   \label{Delta}
\end{equation}
and\footnote{The implicit contractions over indices are always over the
maximal set of indices for the matrices~: $\left(
n^{-1}Nz\right){}^A\equiv \left( n^{-1}\right){}^{AB}N_{BI}z^I$}
\begin{equation}
{\cal M}^{-1\, A\bar B}=\left( n^{-1}\right){}^{AB}+\Delta ^{-1}\left(
n^{-1}Nz\right){}^A \left( n^{-1}N\bar z\right){}^B\end{equation}

Furthermore we have the following result:
if $X^I S_I =0$ then
\begin{eqnarray}
R_I \left( N^{-1} \right){}^{IJ} S_J &=& R_A \left( n^{-1}\right){}^{AB}
S_B \nonumber\\
&&+\Delta ^{-1} \left( -R_0 +R_A(n^{-1}N)^A_{\ 0}\right)
\left( n^{-1} Nz \right){}^B S_B\ .\nonumber\\
&=& R_A \cM^{-1\,\bar AB}S_B
-\Delta ^{-1} R_I\bar z^I
\left( n^{-1} Nz \right){}^B S_B\ .  \label{lemma}
\end{eqnarray}
The last term disappears if moreover $R_I$ satisfies $\bar X^I R_I=0$, a
result which was given already in \cite{BEC}.

\section{Symplectic reparametrizations of special \Ka\ manifolds}
\label{app:symrepar}
\setcounter{equation}{0}

In \cite{CecFerGir} it was noted that two different $F$ functions
can give rise to equivalent theories (provided the vector fields
are abelian). As an example it was demonstrated
that two $n=1$ theories discussed in \cite{BEC} with $F^{\rm
I}=i(X^1)^3/X^0$ and $F^{\rm II}=( X^0)^{1/2}\, ( X^1)^{3/2}$
lead to equivalent field equations. The relation between two
functions that describe equivalent theories takes the form of
symplectic reparametrizations of the same type as the duality
invariances discussed in \cite{dWVP}; the invariances can be
viewed as a special subclass. The space of theories describing
$n$ abelian vector multiplets coupled to $N=2$ supergravity is
thus parametrized by holomorphic functions of $n+1$ variables
that are homogeneous of second degree, divided by $Sp(2n+2,
\Rbar)$ transformations \cite{CecFerGir}.
In addition, functions that differ by a quadratic polynomial with
imaginary coefficients, are equivalent \cite{dWVP}.
The symplectic
reparametrizations are relevant in the treatment of Calabi-Yau
manifolds and related phenomena and for bringing the theory
in a form where certain subgroups of the duality invariances
are linearly realized (see,
for instance, \cite{Cand,FerLusThe,Cecotti,FreSor,Villas}). Here
we summarize their main features and clarify a number of related
issues.

For the scalars the possibility of such reparametrizations is already
suggested by the part of the action depending only on scalars and
gravitons (after elimination of the auxiliary $D$ field, see
\cite{dWLVP}), which
is invariant under local scale and phase transformations, \begin{eqnarray}
4e^{-1}{\cal L}_{(0,2)} &=& \textstyle{\frac{1}{6}}R\,(\bar X^I\, F_I +
X^I\,\bar F_I) \label{DAlagr} \\[1mm]
&&+ (\partial_\mu -iA_\mu)\bar X^I\,(\partial^\mu +iA^\mu)F_I
+ (\partial_\mu +iA_\mu)X^I\,(\partial^\mu -iA^\mu)\bar F_I,
\nonumber\end{eqnarray}
where $A_\mu$ is the (auxiliary) gauge field associated with the
phase transformations.
One may now perform linear redefinitions of $(X^I, F_I)$,
which commute with the local scale and phase transformations
by virtue of the fact that $F(X)$ is holomorphic and homogeneous
of second degree,
\begin{eqnarray}
\tilde X{}^I&=&U^I_{\ J}\,X^J -{\textstyle{1\over2}}i Z^{IJ}\,
F_J,\nonumber\\
\tilde F{}_I&=&V_I{}^J\,F_J +2i W_{IJ}\,X^J ,
\label{xxx}
\end{eqnarray}
where $U$, $V$, $W$ and $Z$ are constant $(n+1)\times(n+1)$
matrices.
The above redefinition preserves the form of
the Lagrangian (\ref{DAlagr})
provided that (for generic functions $F(X)$, so that $X^I$ and
$F_I$ may be considered as linearly independent functions) the matrix
\begin{equation}
{\cal O} =\left(\begin{array}{cc}
U & Z \\[1mm]
W & V
\end{array}\right)  \label{uvzwg}
\end{equation}
satisfies
\begin{equation}
{\cal O}^{-1} = \Omega\, {\cal O}^{\dagger} \,\Omega^{-1}\ ,\quad
\mbox{where} \quad \Omega =\left(\begin{array}{cc}
0&\unity \\[1mm]
-\unity & 0
\end{array}\right) .
\label{spc}
\end{equation}
For functions $F$ for which $F_I$ and $X^I$ are dependent, the
derivation given above is not fully applicable. In that case it is
best to proceed to the subsequent derivation for the spinor and
vector fields, which is more general, and subsequently return to the
scalar-field sector \cite{dWVP,FerLusThe}.

We shall need that $\partial\tilde X{}^I/\partial X^J$ is
non-singular. In other words, that the first line of \eqn{xxx},
with $F_I$ a function of $X$, defines  an
invertable relation between $\tilde X{}^I$ and $X^I$.
This is an additional non-trivial condition on the matrix
\eqn{uvzwg}, which depends on $F$. To see this, consider for instance the
transformation defined by $\tilde X^1=\half i F_1$ and $\tilde
F_1=2iX^1$, with all other variables left unchanged, which satisfies
\eqn{spc}. However, if $F$ depends linearly on $X^1$,
this transformation is not allowed, as $\tilde X^I$ does not
depend on $X^1$, so that the relation between  $\tilde X^I$ and
$X^I$ is not invertable.

In order to remain within the same class of Lagrangians,
one must also require that $\tilde F{}_I$ can be written as the
derivative of some new function $\tilde F(\tilde X)$,
\begin{equation}
\tilde F_I = {\partial \tilde F(\tilde X)\over \partial\tilde
X{}^I} \ ,
\end{equation}
which is again holomorphic and homogeneous of second degree. This
is the case when
\begin{equation}
\tilde F_{IJ}\equiv \frac{\partial \tilde F_I}{\partial \tilde
X^J}=\left(V_I{}^KF_{KL}+2iW_{IL}\right) \frac{\partial
X^L}{\partial \tilde X^J}  \label{F2}
\end{equation}
is a symmetric matrix. Here we made use of the fact that the
transformation $\tilde X^I(X^I)$ is invertable, as explained above. Using
the first line of \eqn{xxx} this integrability condition requires that
\begin{equation}
2i (U^{\rm T} W)_{IJ} + (U^{\rm T}V)_I{}^K\, F_{KJ} +F_{IK}\,
(Z^{\rm T} W)^K{}_J -\textstyle{\frac{i}{2}}F_{IK}\,(Z^{\rm T}
V)^{KL}\,F_{LJ}
\end{equation}
be symmetric in $I$ and $J$.
For a general function $F$ the above condition implies that the
first and the last term are separately symmetric; furthermore we assume
that the identity is the only constant matrix that commutes with
$F_{IJ}$. As a result one finds
\begin{equation}
{\cal O}^{-1} = \Omega\, {\cal O}^{\rm T} \,\Omega^{-1} \ .
\label{spr}
\end{equation}
Note, however, that for special functions this restriction may
not necessarily be required at this point.

Combining \eqn{spr} with \eqn{spc} we conclude that $\cal O$ must
be an element of $Sp(2n+2, \Rbar)$ with unit
determinant\footnote{In view of the local scale and phase
invariance ${\cal O}$ may always be multiplied by an arbitrary
complex function of the space-time coordinates.  We have
suppressed these possible
multiplicative factors by restricting the
determinant of $\cal O$ to unity.}. The new function $\tilde F$
is again holomorphic and homogeneous, and follows directly from
(\ref{xxx}),
\begin{eqnarray}
\tilde F(\tilde X) &=& {\textstyle{1\over2}}\tilde X{}^I \,
\tilde F_I \label{Ftilde} \\[1mm]
& =& i\big(U^{\rm T}W\big)_{IJ}X^I X^J +{\textstyle{1\over2}}
\big(U^{\rm T}V + W^{\rm T}  Z\big)_I{}^J X^IF_J
-{\textstyle{1\over4}}i \big(Z^{\rm T}V\big){}^{IJ}F_I F_J .
\nonumber
\end{eqnarray}
Observe that the above result does not correspond to a
reparametrization of the function $F$ itself, as this would
lead to $\tilde F(\tilde X)=F(X)$. In this connection it is
relevant that
the addition of a quadratic polynomial with imaginary
coefficients to $F(X)$ does not change the action \cite{dWVP}.

We conclude that the Lagrangians \eqn{DAlagr} parametrized by
$F(X)$ and $\tilde F(\tilde X)$ represent equivalent theories.
Furthermore, when
\begin{equation}
\tilde F(\tilde X) = F(\tilde X), \label{Finv}
\end{equation}
the Lagrangian is {\em invariant} under the symplectic
transformations. Note that this does {\em not} imply that $F$
itself is an invariant function. Indeed, from comparison to
\eqn{Ftilde} one readily verifies that $F(\tilde X)\not= F(X)$,
as was already observed in \cite{dWVP} (cf. \eqn{delFd}).

The {\em invariant} reparametrizations, called duality
invariances, were studied
extensively in \cite{dWVP} in the context of infinitesimal
transformations. On the basis of this work it is clear how to
extend the reparametrizations to the other fields. Usually they
cannot be formulated for the abelian
vector fields but are realized on the field strengths by
generalized duality transformations \cite{dual}. Hence the
symplectic parametrizations show that the field equations
corresponding to different functions $F$ are equivalent, but
not necessarily the actions. Before
discussing the symplectic reparametrizations for the other
fields, let us introduce some additional notation. Generic
$(2n+2)$-dimensional vectors that transform under the action of
the $Sp(2n+2,\Rbar)$ will be denoted by $(u^I, v_J)$. Under the
infinitesimal reparametrizations specified by \eqn{Oinf},
the vector $(u^I, v_J)$ thus transforms according to
\begin{eqnarray}
\delta u^I&=&B^I{}_{\!J}\, u^J-D^{IJ}\,v_J\nonumber\\
\delta v_I&=&C_{IJ}\,u^J-B^J{}_{\!I}\, v_J \ .\label{infdual}
\end{eqnarray}
As we have seen, the vector $(X^I,-{1\over 2}iF_J)$ is an example
of such an $Sp(2n+2,\Rbar)$ vector. As we shall discuss below,
the spinor and vector fields lead to similar vectors. The
symplectic invariant of two $Sp(2n+2,\Rbar)$ vectors, $(u^I,
v_J)$ and $(u^{\prime I}, v^\prime_J)$ takes the form
\begin{equation}
\Omega(u,v;u',v') \equiv u^I\,v'_I - v_I\, u^{\prime I},
\label{symplinv}
\end{equation}
and satisfies $\Omega(\tilde u, \tilde v;\tilde u', \tilde v') =
\Omega(u,v;u',v')$.

The transformation of the fields $X^I$ can be written with the aid
of a field-dependent matrix ${\cal S}(X)$ according to (cf.
\eqn{xxx})
\begin{equation}
X^I\to \tilde X^I= {\cal S}^I{}_{\!J}(X)\, X^J,  \label{XcStr}
\end{equation}
where
\begin{equation}
{\cal S}^I{}_{\!J}(X) \equiv U^I{}_{\!J} -{\textstyle{1\over 2}} i
Z^{IK}\,F_{KJ}  \ .  \label{cSfin}
\end{equation}
Note that this result reduces to \eqn{scaltra} for
infinitesimal transformations as
${\cal S}^I{}_{J}\approx \delta ^I_J +
B^I_{\ J}+\frac{1}{2}iD^{IK}F_{KJ}$.
It is not difficult to show that
$\cal S$ satisfies the group  property of $Sp(2n+2,\Rbar)$,
\begin{equation}
{\cal S}_1(\tilde X) \,{\cal S}_2(X)=  {\cal S}_3(X)\ ,
\end{equation}
where ${\cal S}_1$, ${\cal S}_2$ and ${\cal S}_3$ correspond to
the $Sp(2n+2,\Rbar)$ matrices ${\cal O}_1$, ${\cal O}_2$ and
${\cal O}_3$, with ${\cal O}_3= {\cal O}_1 {\cal O}_2$ and
$\tilde X= {\cal S}_2 \,X$.
Here use was made of \eqn{F2}, which can be written as
\begin{equation}
\tilde F_{IJ} = \big(V_I{}^KF_{KL}+2iW_{IL}\big) \,\big({\cal
S}^{-1} \big)^L{}_{\!J} = \big({\cal S}^{-1} \big)^L{}_{\!I}
\big(F_{LK}\, V_J{}^K+2iW_{JL}\big)\ , \label{F2tilde}
\end{equation}
where the right-hand side depends on the coordinates $X^I$
through $F_{IJ}$ and $\cal S$.
{}From \eqn{F2tilde} it is straightforward to derive the following
useful formulae for symplectic reparametrizations of the tensors
$N_{IJ}$ and $F_{IJK}$,
\begin{eqnarray}
\tilde N_{IJ} &=& N_{KL}\, \big(\bar{\cal S}^{-1}\big)^K{}_{\!I}\,
\big({\cal S}^{-1}\big)^L{}_{\!J} , \nonumber\\[1mm]
\tilde F_{IJK} &=& F_{MNP}\, \big({\cal S}^{-1}\big)^M{}_{\!I} \,
\big({\cal S}^{-1}\big)^N{}_{\!J}\,
\big({\cal S}^{-1}\big)^P{}_{\!K} ,\label{NFcStr}
\end{eqnarray}

Given an $Sp(2n+2,\Rbar)$ vector $(u^I, v_J)$, we may construct
(complex) $(n+1)$-component vectors by
\begin{equation}
{\cal V}_I= v_I +{\textstyle\half}{i} F_{IJ}u^J = N_{IJ}{\cal
V}^J ,
\end{equation}
which transform under $Sp(2n+2,\Rbar)$ according to
\begin{equation}
\tilde{\cal V}_I = {\cal V}_J\,\big({\cal S}^{-1}
\big)^J{}_{\!I}\ ,\qquad
\tilde{\cal V}^I = \bar{\cal S}^I{}_{\!J}\,  {\cal V}^J
\ .\label{gencStr}
\end{equation}
On the other hand, given a vector ${\cal V}^I$ one constructs an
$Sp(2n+2,\Rbar)$ vector by
\begin{equation}
\big(u^{\prime I}, v'_J\big) = \big({\cal V}^I,
{\textstyle{1\over 2}} i\bar F_{JK}\,{\cal V}^K\big) \ ,
\end{equation}
whose imaginary part equals the original vector $(u^I,v_J)$
provided the latter were real.

We now turn to the reparametrizations of the spinor
fields $\Omega^I_i$ and $\Omega^{iI}$ (the index
$i$ refers to chiral $SU(2)$; an upper (lower) index denotes the
positive (negative) chirality components). Their reparametrizations
can be inferred from the fact that both
\begin{equation}
\bar X^I\, N_{IJ}\,\Omega^J_i =\bar X^I\,\big(F_{IJ}\,
\Omega^J_i\big) + \bar F_I\, \Omega^I_i
\end{equation}
and the kinetic term of the spinors and their coupling to the
scalars \cite{dWLVP},
\begin{equation}
e^{-1} {\cal L}^{\rm fermionic} =  {\textstyle{1\over
16}}\left\{\bar \Omega^{iI}
\stackrel{\leftrightarrow}{\rlap{D}/}\big(F_{IJ}\Omega_i^J\big) +
\big(\bar F_{IJ} \Omega^{iJ}\big)
\stackrel{\leftrightarrow}{\rlap{D}/} \Omega^{I}_i
\right\} \ ,
\end{equation}
should preserve their form under symplectic reparametrizations.
Both these expressions take the form of the symplectic
invariant \eqn{symplinv} provided we identify
$(\Omega_i^I,-{1\over2}iF_{JK}\,\Omega_i^K)$ as an $Sp(2n+2,
\Rbar)$ vector. This is equivalent to the reparametrization
\begin{equation}
\tilde \Omega^I_i = {\cal S}^I{}_{\!J}\, \Omega_i^J \ .
\end{equation}

What remains are the reparametrizations for the vector fields,
which, as emphasized above, are realized on the field strengths.
First we evaluate the tensor $\cN$ corresponding to the function
$\tilde F(\tilde X)$. Denoting this tensor by
$\tilde\cN$, one finds
\begin{equation}
\tilde \cN_{IK} \big(U^K{}_{\!J}+2iZ^{KL}\,\cN_{LJ} \big) =
-{\textstyle{1\over2}}i W_{IJ} +V_I{}^{\!K}\, \cN_{KJ} \ .
\label{Ntilde}
\end{equation}
As a consistency check, one may multiply both sides of
this equation by $\tilde X^I$ and $X^J$; then both
sides become proportional to $\tilde F(\tilde X)$ upon using
\eqn{xxx} and the identities $\cN_{IJ}\,X^J=- {1\over 4}F_I$ and
$F_I\, X^I= 2 F$.

On the other hand, on the basis of the
results of \cite{dWVP},
we expect that the self-dual component of the abelian field
strength gives rise to an $Sp(2n+2,\Rbar)$ vector $(F^{+
I}_{\mu\nu}, 2i\cN_{JK}\,F^{+ K}_{\mu\nu})$. This implies that
\begin{equation}
2i\tilde {\cal N}_{IJ}\,\tilde F_{\mu \nu }^{+I}=2i\tilde {\cal
N}_{IJ} \left( U +2iZ{\cal N}\right)^J{}_{\!K}\, F_{\mu \nu
}^{+K} =2i\left( V{\cal N} +W\right)_{IJ}\, F_{\mu \nu}^{+J}\ .
\end{equation}
which is precisely in accord with the equation \eqn{Ntilde} found
above.

After the reduction of the four-dimensional fields we obtain new
vectors $W^I_\mu$, $A^I$ and $B_I$, which are related to the
field strengths. From this it follows that $\big(W^I_\mu ,
2i\cN_{JK}\, W_\mu^K\big)$ and $\big(A^I,B_J\big)$
are also $Sp(2n+2, \Rbar)$ covariant vectors.

The above expressions for the symplectic reparametrizations were
all derived and/or used in \cite{dWVP,dWVP2} in infinitesimal
form for reparametrizations that constitute an
invariance of the theory, i.e., reparametrizations that satisfy
\eqn{Finv}. What we
want to stress here is that the restriction to invariance
transformations is not necessary. Exactly the same form of the
transformations is obtained for arbitrary symplectic
reparametrizations.

Finally, when the duality invariances act transitively on the
manifold of the spinless fields, transformation rules such as the
ones derived above lead to a complete determination of
the various functions of the spinless fields that appear in the
supergravity Lagrangian, a feature which was extensively
exploited in the explicit construction of many supergravity
Lagrangians in the past.
\vspace{0.4cm}

For the convenience of the reader we list the
$(2n+2)$-component objects transforming
under the duality transformations as in \eqn{xxx},
or infinitesimally as in \eqn{infdual}, that we encountered in
this paper; they are
\begin{equation}
(X^I, -\ft12 iF_J)\ , \ \
(F^{+ I}_{\mu\nu}, 2i\cN_{JK}\,F^{+ K}_{\mu\nu})\ , \ \
(A^I,B_I)\ , \ \
(W ^I, 2i\cN _{IJ}W ^J)\ ,
\end{equation}
and their complex conjugates.
Furthermore we encountered quantities transforming according to
\eqn{cSfin},  such as $X^I$ (cf. \eqn{XcStr}),
the tensors transforming as in \eqn{NFcStr}, and
$\cB_I$ and $\cB^I$, defined in \eqn{defcB}, which
transform as ${\cal V}_I$ or ${\cal V}^I$, respectively (cf.
\eqn{gencStr}).

\end{document}